\documentclass[12pt,journal,cspaper,onecolumn,compsoc]{IEEEtran}
\pdfoutput=1
\newenvironment{ItemizeStandard}{\begin{itemize}%
\addtolength{\parsep}{8mm}
\addtolength{\topsep}{0.75mm}
\addtolength{\itemsep}{0.75mm}}
{\end{itemize}}

\newenvironment{EnumerateRoman}{\begin{enumerate}%
\setlength{\parsep}{0pt}\setlength{\topsep}{0.1\baselineskip}\setlength{\itemsep}{0.1\baselineskip}}{\end{enumerate}}

\usepackage{amsmath,amsfonts,amssymb}
\allowdisplaybreaks
\usepackage{mathrsfs} 
\usepackage{units}
\usepackage{multirow}
\newcommand{\FigScaleMultipleTwo}{0.565}
\newcommand{\FigScaleMultipleFour}{0.55}    
\newcommand{\FigScaleMultipleFourB}{0.5522}   
\newcommand{\SinglePlotScale}{0.65}
\usepackage{graphicx} 
\usepackage[tight,footnotesize]{subfigure} 
\usepackage{array} 
\usepackage[footnotesize]{caption} 
\usepackage{tabularx}
\usepackage{ctable}

\usepackage{url}

\usepackage[nocompress]{cite} 

\DeclareMathOperator{\oD}{D}                                              
\DeclareMathOperator{\oE}{E}                                               
\DeclareMathOperator{\oH}{H}                                              
\DeclareMathOperator{\oP}{P}                                               
\newcommand{\cR}{\mathcal{R}}                                          
\newcommand{\cP}{\mathcal{P}}                                           

\newcommand{\phicrit}{\varphi_\textnormal{crit}}

\newtheorem{proposition}{Proposition}

\newtheorem{theorem}[proposition]{Theorem}
\newtheorem{lemma}[proposition]{Lemma}
\newtheorem{corollary}[proposition]{Corollary}
\newcommand{\BeginProof}{\noindent\hspace{2em}{\itshape Proof: }}
\newcommand{\ProofSquare}{\blacksquare}
\newcommand{\EndProof}{\hspace*{\fill}~$\ProofSquare$\par\unskip}
\newcommand{\SpaceAfterPropositionEndedWithFormula}{\vspace{\belowdisplayskip}}

\begin{document}

\title{Smart Deferral of Messages for\\ Privacy Protection in Online Social Networks}
\author{Javier~Parra-Arnau, F\'elix~G\'omez~M\'armol, David~Rebollo-Monedero and~Jordi~Forn\'e
\IEEEcompsocitemizethanks{\IEEEcompsocthanksitem J. Parra-Arnau, D. Rebollo-Monedero and J. Forn\'e and are with
the Department of Telematics Engineering, Universitat Polit\`ecnica de Catalunya,
C.\ Jordi Girona 1-3, E-08034 Barcelona, Spain.\protect\\
E-mail: {javier.parra,david.rebollo,jforne}@entel.upc.edu.
\IEEEcompsocthanksitem F. G\'omez M\'armol is with NEC Laboratories Europe, Kurf\"ursten-Anlage 36, 69115 Heidelberg, Germany.\protect\\
E-mail: felix.gomez-marmol@neclabs.eu.}
\thanks{Manuscript prepared January, 2014.}}


\IEEEcompsoctitleabstractindextext{%
\begin{abstract}
Despite the several advantages commonly attributed to social networks such as easiness and immediacy to communicate with acquaintances and friends, significant privacy threats provoked by unexperienced or even irresponsible users recklessly publishing sensitive material are also noticeable. Yet, a different, but equally hazardous privacy risk might arise from social networks profiling the online activity of their users based on the timestamp of the interactions between the former and the latter. In order to thwart this last type of commonly neglected attacks, this paper presents a novel, smart deferral mechanism for messages in online social networks. Such solution suggests intelligently delaying certain messages posted by end users in social networks in a way that the observed online-activity profile generated by the attacker does not reveal any time-based sensitive information. Conducted experiments as well as a proposed architecture implementing this approach demonstrate the suitability and feasibility of our mechanism.
\end{abstract}

}

\maketitle

\IEEEdisplaynotcompsoctitleabstractindextext

%
\IEEEpeerreviewmaketitle

\setcounter{footnote}{0} 

\section{Introduction}
\label{sec:Introduction}
\IEEEPARstart{I}{nformation} and communication technologies (ICT) have revolutionized our lives, leading to an unprecedented societal transformation aimed to reach the so called ``digital era''. In that sense, we are witnessing today how social networks are paving the way to reach such transformation by influencing and even modifying the way we interact with each other and behave amongst us. Amid the plethora of advantages brought by social networks we find the easiness to communicate with friends and acquaintances, the easiness to share thoughts, opinions and experiences in any format (plain text, pictures, audio, video, etc.) and even the immediate reaction in case of emergency or catastrophe.

Yet, despite their proven convenience, online social networks might also pose non-negligible privacy risks~\cite{GomezMarmolIC14}, most of the time due to irresponsible or unexperienced users who recklessly post private or sensitive information exposing themselves (and sometimes maybe even their friends and connections in the social network) to undesired and unexpected situations (bullying, bribery, identity theft, etc.).

In this manuscript, however, we will focus on a different privacy threat inherent to social networks. Although equally hazardous, such threat is not based on the content itself published by the end users and, therefore, it might not be as evident as the latter. Whenever we interact with any social network (post a comment on Facebook, write a message in Twitter, etc.), regardless the content (or lack of it) associated with such interaction, it is reasonably easy for the social network to log the timestamp when the interaction occurred. By doing so, the social network is able to build, almost effortlessly, an activity profile of its users based on the timestamps of each of the interactions conducted by such users within the social network.

It is worth noting the significance and danger that such a time profiling threat entails. Some examples illustrating the kind of sensitive, private information that could be inferred by social networks from a profile of online activity include, for instance: whether a particular user is unemployed or not, when this user seems to be at home, whether this user is single or married, when this user normally wakes up and goes to bed, whether this user is on holidays or not, etc.

Another two reasons making this risk particularly perilous are: on the one hand, the fact that the type of attacker who could perform this kind of profiling would be a rudimentary adversary who does not possess enough resources to analyze the content of the information and therefore has to resort to timing information. And, on the other hand, the fact that encrypting the content is useless in this scenario. The encryption of users' data can prevent a privacy attacker from profiling users based on the content of the information revealed, but clearly not against an adversary who relies on timing information. In other words, the attacker does not need to decrypt messages to infer user activity.

With the aim of hindering this kind of attack, the paper at hand presents a work where a novel, smart deferral mechanism is investigated. The mechanism under study enables users to delay of a number of their messages (without loss of generality, interactions with social networks) incapacitating these social networks to break their privacy and snoop in their habits by creating online activity profiles as described before. As a consequence, the observed profile generated by the attacker (which in our case, as we will see later, is not limited to the social networking site, but broadened to any entity able to collect such timing information), will differ from the original, genuine user profile of online activity in such a way that the attacker is unable to either i) individuate users (find users whose profiles significantly deviate from the average population), nor ii) classify them (categorize users within a particular group of users based on their activity profiles).

The remainder of the paper is organized as follows: Sec.~\ref{sec:StateOfTheArt} analyzes some remarkable works found in the literature within the context of privacy-enhancing technologies (PETs). Our smart deferral mechanism is introduced and described in Sec.~\ref{sec:Technique}, while Sec.~\ref{sec:Architecture} specifies the building blocks of an architecture implementing our solution. In turn, Sec.~\ref{sec:DelayCapacity} studies two specific utility metrics for our approach, namely, expected message delay and messages storage capacity. A comprehensive set of experiments demonstrating the feasibility of our proposal has been conducted and its outcomes are shown in Sec.~\ref{sec:Experiments}. Finally, Sec.~\ref{sec:Conclusion} presents some concluding remarks as well as future research directions.

\section{State of the Art}
\label{sec:StateOfTheArt}
\noindent
To the best of our knowledge, there is no privacy-enhancing mechanism \emph{specifically conceived} to counter the time-based profiling attack described in Sec.~\ref{sec:Introduction}. In this section, we review some general-purpose technologies that could be used to cope with this kind of attacks. Partly inspired by~\cite{Shen07SIGIR}, we classify these technologies into four categories: encryption-based methods, approaches based on trusted third parties (TTPs), collaborative mechanisms and data-perturbative techniques.

In traditional approaches to privacy, users or designers decide whether certain sensitive information is to be made available or not. On the one hand, the availability of this data enables certain functionality, e.g., sharing pictures with friends on a social network. On the other hand, its unavailability, traditionally attained by means of access control or encryption, produces the highest level of privacy. In the scenario considered in this work, the use of encryption-based techniques could limit access to the content of the messages posted on a social network, by providing or not a cryptographic key permitting their deciphering. Nevertheless, even though this key was not provided, an attacker with access to the encrypted messages could still be able to jeopardize user privacy --- encryption may conceal the content of such messages, but it cannot hide the time instants when they were posted.

A conceptually-simple approach to protect user privacy consists in a TTP acting as an intermediary or \emph{anonymizer} between the user and an untrusted information system. In this scenario, the system cannot know the user ID, but merely the identity of the TTP itself involved in the communication. Alternatively, the TTP may act as a \emph{pseudonymizer} by supplying a pseudonym ID' to the service provider, but only the TTP knows the correspondence between the pseudonym ID' and the actual user ID. In online social networks, the use of either approach would be unappropriated as users of these networks are required to be logged in. Although the adoption of TTPs to this end would therefore be ruled out, users themselves could provide a pseudonym at the sign-up process, thus playing the role of a pseudonymizer. In this line, some sites have started offering social-networking services where users are not required to reveal their real identifiers\footnote{SocialNumber (\url{http://www.socialnumber.com}) is an example of such networks, where users must choose a unique number as identifier.}.

Unfortunately, none of these approaches may prevent an attacker from profiling a user based on message content, and ultimately inferring their real identity. In its simplest form, reidentification is possible due to the personally identifiable information often included in the messages posted. However, even though no identifying information is included, pseudonyms could also be insufficient to protect both anonymity and privacy. As an example, suppose that an observer has access to certain behavioral patterns of online activity associated with a user, who occasionally discloses their ID, possibly during interactions not involving sensitive data. The same user could attempt to hide under a pseudonym ID' to exchange information of confidential nature. Nevertheless, if the user exhibited similar behavioral patterns, the unlinkability between ID and ID' could be compromised through these similar patterns. In this case, any past profiling inferences carried out for the pseudonym ID' would be linked to the actual user ID.

Another class of PETs relying on trusted entities is anonymous-communication systems (ACSs). In anonymous communications, one of the goals is to conceal who talks to whom against an adversary who observes the inputs and outputs of the anonymous communication channel. Mix systems~\cite{Chaum81CACM,Cottrell94,Danezis03PET} are a basic building block for implementing anonymous-communication channels. These systems perform cryptographic operations on messages such that it is not possible to correlate their inputs and outputs based on their bit patterns. In addition, mixes delay and reorder messages to hinder the linking of inputs and outputs based on timing information.

In the context of our work, ACSs may hide the link between social networking sites and users, and therefore may protect user privacy against the intermediary entities enabling the communications between them. We may distinguish between two cases --- the case where messages are public, and the case where messages are kept private or available to authorized users. In the former case, ACSs obviously cannot provide any privacy guarantees, as user online activity is publicly available. In the latter case, the use of anonymous communications might contribute to privacy enhancement provided that the attacker is an external entity\footnote{Clearly, if the attacker was the social networking platform, any information disclosed by the user would be known to the adversary.}. However, since the adversary model assumed in ACSs considers that the attacker knows all the senders (inputs) and receivers (outputs), it would be enough for this attacker to observe the messages generated by the target user. In short, anonymous communications may not be an appropriate approach to thwart an adversary who strives to profile users based on their online activity.

A particularly rich group of PETs are those where users collaborate to protect their privacy. One of the most popular is \emph{Crowds}~\cite{Reiter98ISS}, which contemplates that a group of users wanting to browse the Web will collaborate to submit their requests. With this purpose, a user wishing to send a request to a Web server selects first a member of the group at random, and then forwards the request to it. When this member receives the request, it flips a biased coin to determine whether to send the request to another member or to submit it directly to the Web server. This process is repeated until the request is finally relayed to the intended destination. As a result of this probabilistic protocol, the Web server and any of the members forwarding the request cannot ascertain the identity of the true sender, that is, the member who initiated the request.

While Crowds and similar collaborative protocols~\cite{Domingo09DKE,Rebollo09COMCOM,Domingo12INS} may be effective in applications such as information retrieval and Web search, the fact is that they are not suitable for the application at hand. The main reason is that users are required to be logged into online social networks. That is, users participating in a collaborative protocol would need the credentials of their peers to log in, which in practice would be unacceptable. Besides, even though users were willing to share their credentials, this would not avoid profiling based on the observation of the messages posted on the social network in question.

An alternative to hinder an attacker in its efforts to profile users consists in perturbing the information they disclose when communicating with an information system. The submission of false data, together with the user's genuine data, is an illustrative example of data-perturbative mechanism. In the context of information retrieval, query forgery~\cite{Rebollo10IT} prevents privacy attackers from profiling users accurately based on the \emph{content} of queries, without having to trust neither the service provider nor the network operator, but obviously at the cost of traffic overhead.

Clearly, the perturbation of user profiles for privacy protection may be carried out not only by means of the insertion of bogus activity, but also by suppression. An example of this latter kind of perturbation may be found in~\cite{Parra12DKE,Parra12TKDE,Parra10TB}, where the authors propose the elimination of tags as a privacy-enhancing strategy in collaborative-tagging applications. Tag suppression allows users to enhance their privacy to a certain degree, but it comes at the expense of degrading the semantic functionality of those applications, as tags have the purpose of associating meaning with resources. The combination of both strategies, that is, forgery and suppression, is studied in the scenario of personalized recommendation systems~\cite{Parra13PhD,Parra11DPM}. The simultaneous use of these two strategies enables users to submit false ratings to items that do not reflect their preferences, and/or refrain from rating certain items they have an opinion on.

The data-perturbative mechanisms described above aim to prevent an attacker from profiling users based on their \emph{interests}. Although these mechanisms could also be used to avoid profiling attacks based on the \emph{time instants} when users communicate through social networks, we believe that they would not be adopted in practice --- users of social networks would be reticent to eliminate their comments and to generate fake comments, as these actions would have a significant impact on the information-exchange functionality provided by social networks.

\section{Privacy Protection via Message Deferral}
\label{sec:Technique}
\noindent
In the introductory section we emphasized the risk of profiling based on the time instants when users submit messages to a social networking site. In particular, we mentioned that, building on this online behavior, an adversary could extract an accurate snapshot of their profiles of activity throughout time and thus could compromise user privacy.

\begin{figure}
\centering
\includegraphics[scale=0.50]{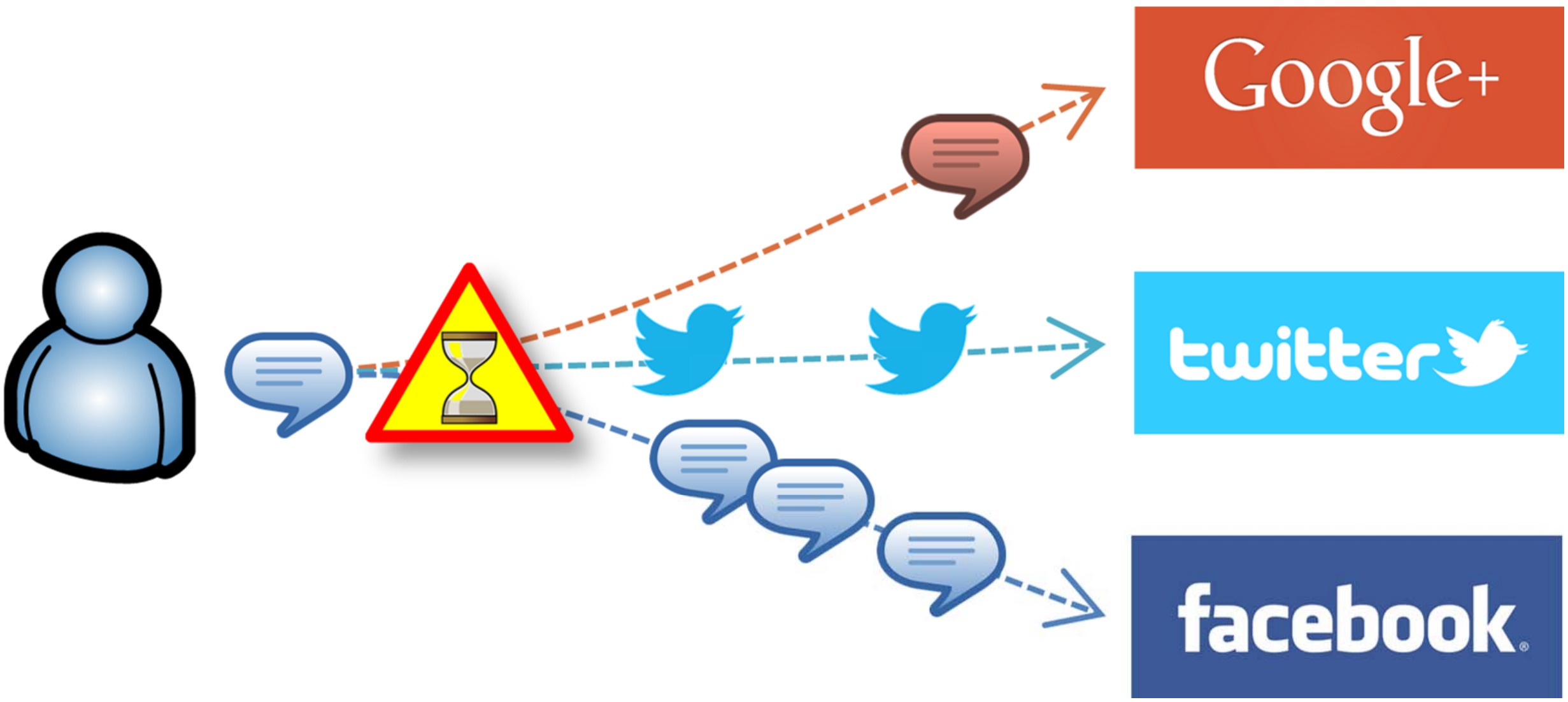}
\caption{Message deferral as a mechanism to protect the privacy of the online activity of a user by delaying the submission of certain messages.}
\label{fig1}
\end{figure}

In this situation, we propose the \emph{deferral of messages} as a conceptually-simple mechanism that may thwart this kind of profiling attacks. The proposed mechanism allows users to delay the submission of certain messages, by storing them locally and afterwards sending them to the social network provider in question. The application of this mechanism may help users to protect their privacy to a certain extent, at the cost of no infrastructure, and without having to trust neither the service provider nor any external entity. The downside of delaying messages, however, is that it comes at the expense of data-storage capacity and, more importantly, the utility of the services provided by the online social network. As an example, consider a user posting a tweet\footnote{A tweet is a message sent using Twitter.} to confirm a meeting this evening. If this tweet was postponed, the confirmation could arrive late and, if so, the information-exchange functionality would be useless. In short, the deferral of messages poses a trade-off between the contrasting aspects of privacy on the one hand, and utility on the other. Fig.~\ref{fig1} illustrates our mechanism.

In the coming sections, we shall investigate the deferral of messages as a technique that may preserve user privacy against an attacker who tries to profile users based on their posting times. Note that this is in contrast to other types of profiling attacks that exploit the \emph{content} of the information disclosed, rather than the \emph{time} when this information is revealed. Obviously, this latter kind of user profiling may, in practice, occur in conjunction with the former.

Although the proposed privacy-enhancing mechanism is explored here in the context of online social networking services, we would like to stress its applicability to other contexts where the submission of data to an information system may enable an adversary to construct a precise reflection of the online activity of a user. The deferral strategy could be used, for instance, in Web search, information retrieval, resources tagging and recommendation systems.

\subsection{Adversary Model}
\label{sec:Technique:Adversary}
\noindent
In order to evaluate the level of privacy provided by our mechanism, it is fundamental to specify the concrete assumptions about the attacker, that is, its capabilities, properties or powers. This is known as the \emph{adversary model} and its importance lies in the fact that the level of privacy provided is measured with respect to it.

Next, we describe the adversary model assumed in this work, in terms of (1) the application scenario considered, (2) the type of adversaries able to profile users, (3) the way these adversaries model user activity, and (4) the objective behind the construction of these activity models.

\begin{figure*}[tb]%
\centering\hspace*{\fill}
\subfigure[]%
{\includegraphics[scale=\FigScaleMultipleTwo]{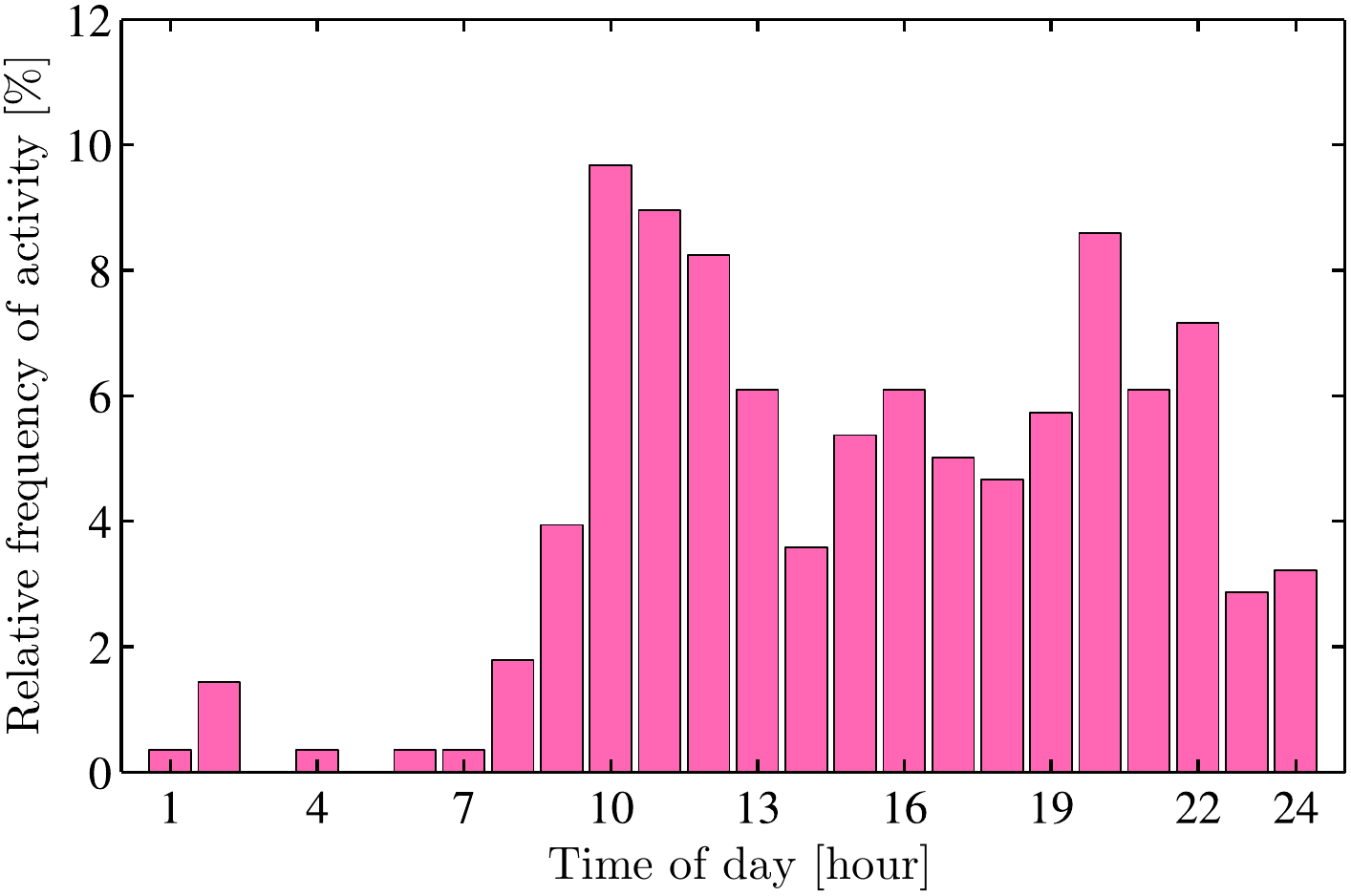}%
\label{fig:chp7:deltarho}}\hfill
\subfigure[]%
{\includegraphics[scale=\FigScaleMultipleTwo]{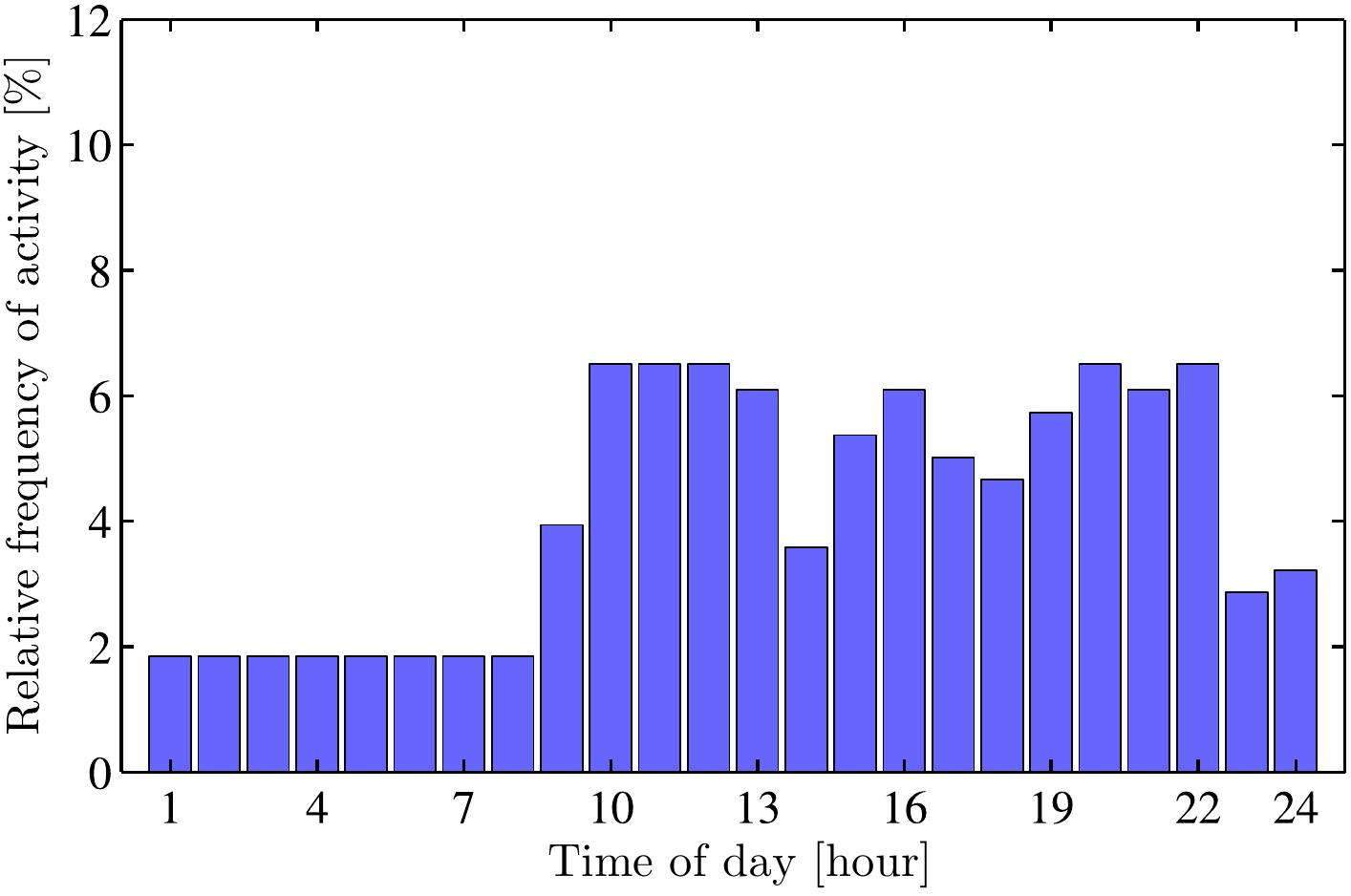}%
\label{fig:chp7:deltasigma}}\hspace*{\fill}
\caption{Actual user profile (a) and apparent user profile (b). Both profiles represent the profile of activity across a day, in particular, the percentage of messages posted between 0 a.m.\ and 1 a.m., 1 a.m.\ and 2 a.m., and so on.}
\label{fig2}
\end{figure*}

\begin{ItemizeStandard}
  \item \textbf{Scenario}. First, we consider a common scenario where users are required to be logged into a social networking site for their messages to be posted. This could be the case of Google Plus, Twitter and Facebook. In addition, we may reasonably assume that users of these applications provide their real identifiers to create their accounts. We must hasten to stress that, even though a user employs pseudonyms, the content of the messages exchanged or the knowledge of their ``friends'' in those social networks may lead an attacker to reidentify this user.

  \item \textbf{Privacy attackers}. In this scenario, any entity capable of capturing users' messages is regarded as a potential privacy attacker. This includes the social network provider, the Internet service provider (ISP), and the intermediary entities (switches, routers, firewalls) enabling the communications between users and social networking sites. Besides, since posted messages are often publicly available\footnote{Messages exchanged on Twitter are publicly visible by default.}, any entity able to collect this information is also taken into consideration in our adversary model.

  \item \textbf{User-profile model}. We assume that the attacker represents behavioral patterns of online user activity as probability mass functions (PMFs). Conceptually, a user profile may be interpreted as a histogram of relative frequencies of messages across a day, week, month or year. The proposed user-profile model is a natural, intuitive representation in line with the models used in many information systems to characterize user profiles~\cite{Xu07WWW,Ye09ICCSE,Erola11SS,Parra13PhD,Parra12DKE,Parra13FGCS}.
\vspace{0.75mm}

        In our adversary model, we distinguish between two kinds of profiles. On the one hand, the user's genuine profile, and on the other, the profile perceived from the outside, which results from delaying certain messages before posting them. Hereafter, we shall refer to these two profiles as the \emph{actual} profile $q$ and the \emph{apparent} profile $t$. That said, in this work we shall assume that the attacker is unaware or ignores the fact that the observed, perturbed profile does not reflect the actual behavior of the user. Fig.~\ref{fig2} provides an example of such profiles. In this figure we represent the profile of online activity of a user within 1-hour slot throughout one day.

 \item \textbf{Objective behind profiling}. Finally, our adversary model contemplates what the attacker is after when profiling users. According to~\cite{Parra13FGCS}, and in line with the technical literature of profiling~\cite{Hildebrandt05FIDIS,Hildebrandt08B}, we consider two possible objectives for the attacker. On the one hand, we may assume the attacker strives to target users who deviate from the average behavior. This is known as \emph{individuation}, meaning that the adversary aims at discriminating a given user from the whole population of users, or said otherwise, wishes to learn what distinguishes that user from the other users. On the other hand, we may consider that the attacker's goal is to classify a user into a predefined group of users.  To conduct this \emph{classification}, the attacker contrasts the user's profile with the profile representative of a particular group.
\end{ItemizeStandard}
These two objectives, together with the assumptions about the scenario and the user profile representation, constitute the adversary model upon which our privacy metric builds.

\subsection{Privacy Metric of Online Activity against Individuation}
\label{sec:Technique:Metric}
\noindent
Next, we justify the Shannon entropy and the Kullback-Leibler (KL) divergence as measures of privacy when an attacker aims to individuate users based on their profiles of activity. The rationale behind the use of these two information-theoretic quantities as privacy metrics is documented in greater detail in~\cite{Parra13FGCS}.

Recall that Shannon's entropy $\oH(t)$ of a discrete random variable (r.v.) with PMF $t=(t_i)_{i=1}^n$ on the alphabet $\{1,\ldots,n\}$ is a measure of the uncertainty of the outcome of this r.v., defined as
\begin{equation*}
\oH(t)=-\sum t_i   \log t_i.
\end{equation*}
Throughout this work, all logarithms are taken to base 2, and subsequently the entropy units are \emph{bits}. Given two probability distributions $t$ and $p$ over the same alphabet, the KL divergence is defined as
\begin{equation*}
\oD(t\,\|\,p)=\sum t_i   \log \frac{t_i}{p_i}.
\end{equation*}

The KL divergence is often referred to as \emph{relative entropy}, as it may be regarded as a generalization of the Shannon entropy of a distribution, relative to another. Conversely, Shannon's entropy is a special case of KL divergence, as for a uniform distribution $u$ on a finite alphabet of cardinality $n$,
\begin{equation}\label{eq1}
\oD(t\,\|\,u)= \log n - \oH(t).
\end{equation}

Leveraging on a celebrated information-theoretic rationale by Jaynes~\cite{Jaynes82P}, the Shannon entropy of an apparent user profile, modeled as a PMF, may be regarded as a measure of privacy, or more accurately, anonymity. The leading idea is that the method of types~\cite{Cover06B} from information theory establishes an approximate monotonic relationship between the likelihood of a PMF in a stochastic system and its entropy. Loosely speaking, the higher the entropy of a profile, the more likely it is, and the more users behave according to it. Under this interpretation, entropy is a measure of anonymity, \emph{not} in the sense that the user's identity remains unknown, but only in the sense that higher likelihood of an apparent profile, believed by an external observer to be the actual profile, makes that profile more common, hopefully helping the user go unnoticed, less interesting to an attacker whose objective is to target peculiar users.

If an aggregated histogram of the population were available as a reference profile $p$, the extension of Jaynes' argument to relative entropy would also give an acceptable measure of anonymity. Recall that KL divergence is a measure of discrepancy between probability distributions, which includes Shannon's entropy as the special case when the reference distribution is uniform. Conceptually, a lower KL divergence hides discrepancies with respect to a reference profile, say the population's, and there also exists a monotonic relationship between the likelihood of a distribution and its divergence with respect to the reference distribution of choice, which enables us to deem KL divergence as a measure of anonymity in a sense entirely analogous to the above mentioned.

Under this interpretation, the Shannon entropy is therefore interpreted as an indicator of the commonness of similar profiles. As such, we should hasten to stress that the Shannon entropy is a measure of \emph{anonymity} rather than privacy, in the sense that the obfuscated information is the uniqueness of the profile behind the online activity, rather than the actual profile itself.

\subsection{Formulation of the Trade-Off between Privacy and Message-Deferral Rate}
\label{sec:Technique:Formulation}
\noindent
In this section, we present a formulation of the optimal privacy-utility trade-off posed by our message-deferral mechanism.

In our mathematical model, we represent the messages of a user as a sequence of independent and identically distributed (i.i.d.) r.v.'s taking on values in a common finite alphabet of $n$ time periods, namely the set $\{1,\ldots,n\}$ for some integer $n\geqslant 2$. As an example, the set of time periods could be the hours of a day or a week, or the days of a month. According to this model, we characterize the \emph{actual} profile of a user as the common PMF of these r.v.'s, $q=(q_1,\ldots,q_n)$. In conceptual terms, our model of user profile is a normalized histogram of messages over those time periods.

Based on this model, we quantify the \emph{initial} privacy level as the Shannon entropy of the user's actual profile, $\oH(q)$. For the sake of tractability, we measure utility as the \emph{deferral rate} $\varphi \in [0,1)$, that is, the ratio of messages to total number of messages that a user is willing to delay.

When a user accepts delaying their tweets, comments or, in general, messages, their actual profile $q$ is seen from the outside as the apparent profile $t=q-s+r$, according to a \emph{storing strategy}~$s$ and a \emph{forwarding strategy}~$r$. These strategies are two $n$-tuples that would tell the user when to retain those messages and when to release them. More specifically, the $i$-th component of the storing strategy is the fraction of messages that this user should store at time period $i$. Similarly, $r_i$ is the proportion of messages to total number of messages that the user should forward at time $i$. Clearly, these two strategies must satisfy that  $s_i,r_i \geqslant 0$, $q_i - s_i + r_i \geqslant 0$, for all $i$, and that $\sum s_i=\sum r_i = \varphi$ so that $t$ is a PMF.

According to this notation, we denote by $\oH(t)$ the (\emph{final}) privacy level and define the \emph{privacy-deferral function} as
\begin{equation}
\label{eq2}
\cP(\varphi)=  \max_{\substack{r,s\\r_i \geqslant 0, \, s_i \geqslant 0,\\ q_i - s_i + r_i\geqslant 0,\\
\sum s_i  = \sum r_i = \varphi}} \oH(q-s+r),
\end{equation}
which models the optimal trade-off between privacy and message-deferral rate.

The optimization problem inherent in this definition belongs to the extensively studied class of convex optimization problems~\cite{Boyd04B}. Most of these problems do not have an analytical solution and thus need to be solved numerically. For this, there exist a number of extremely efficient methods, such as interior-point algorithms. The problem formulated here, however, turns out to be a particular case of a more general optimization problem, for which interestingly there is an explicit closed-form solution, albeit piecewise~\cite[\S 7]{Parra13PhD}.

In practice, this means that we shall be able to find an analytical expression for the \emph{optimal} storing and forwarding strategies, i.e., those strategies that maximize user privacy for a given~$\varphi$. Later on, in Sec.~\ref{sec:DelayCapacity:Preliminaries}, we shall show that~\eqref{eq2} is a particularization of this latter problem.

\begin{figure*}[t!]%
\centering\hspace*{\fill}
\subfigure[]%
{\includegraphics[scale=\FigScaleMultipleTwo]{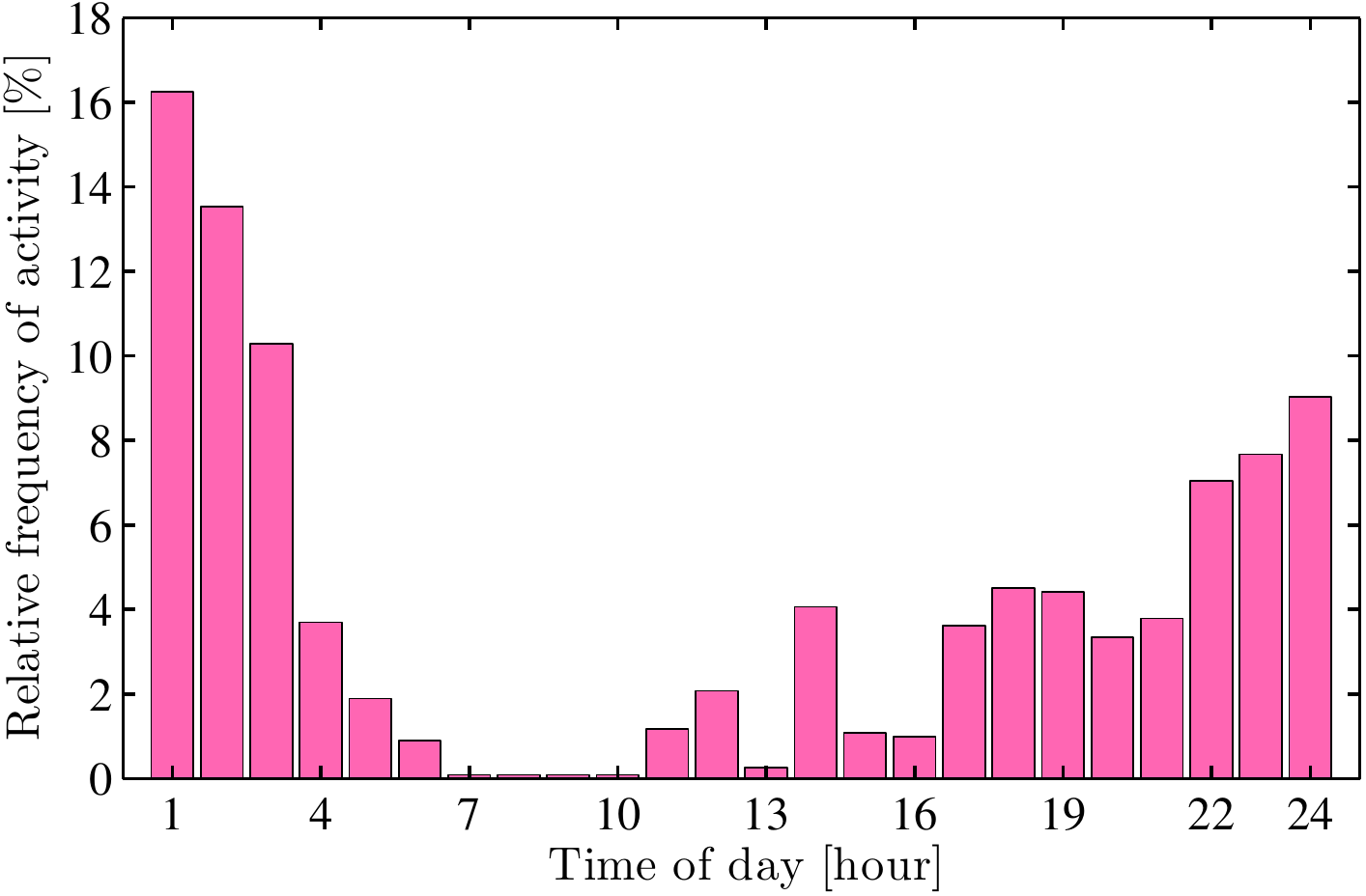}%
\label{fig:chp7:deltarho}}\hfill
\subfigure[]%
{\includegraphics[scale=\FigScaleMultipleTwo]{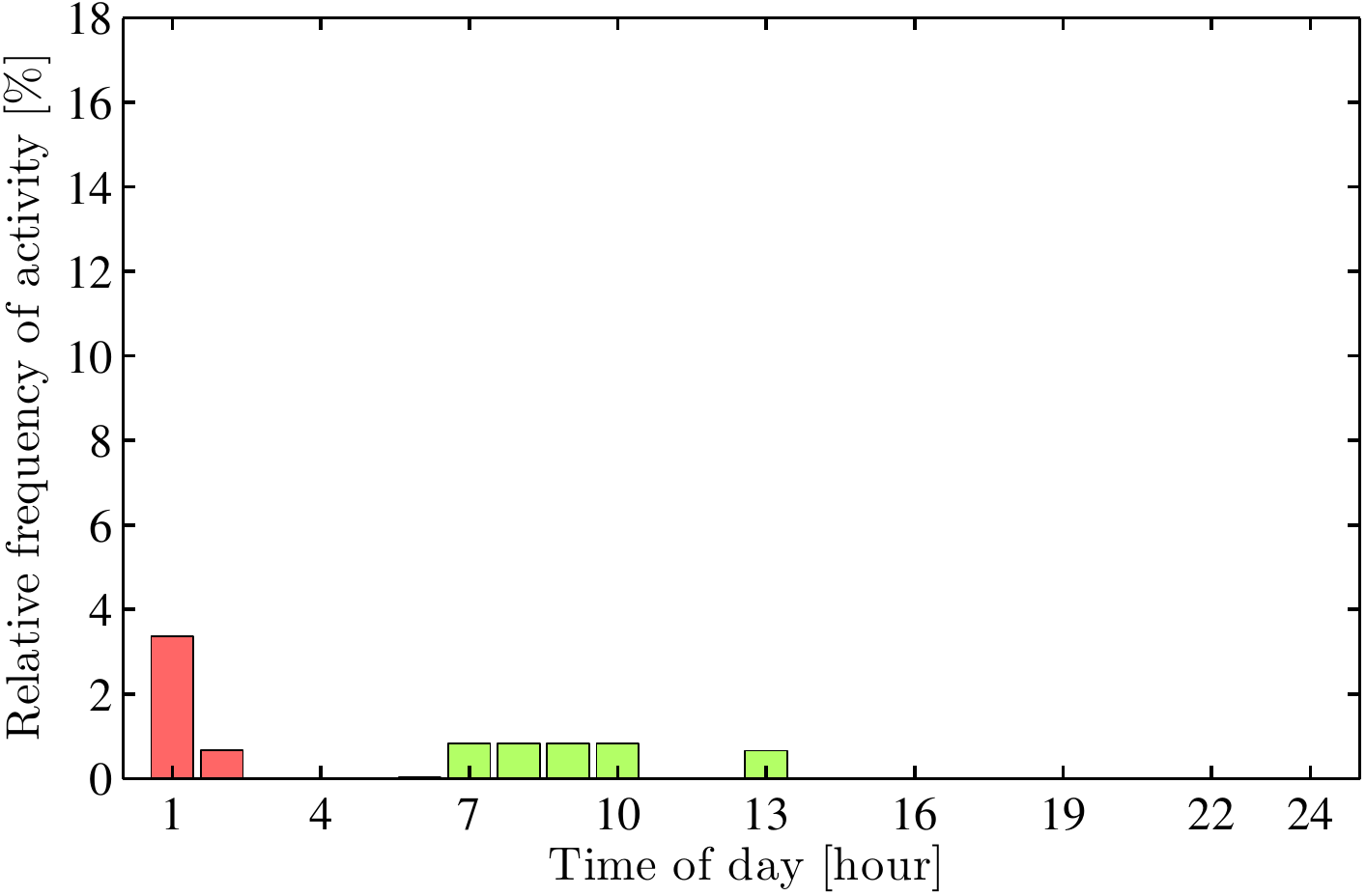}%
\label{fig:chp7:deltasigma}}\hspace*{\fill}
\caption{Example of user profile (a) and its optimal storing and forwarding strategies $r$ and $s$ (b) for a deferral-message rate $\varphi=4\%$.}
\label{fig3}
\end{figure*}

\section{Architecture}
\label{sec:Architecture}
\noindent
In this section we specify the building blocks of an architecture implementing our privacy-enhancing, message-deferral mechanism. The proposed architecture provides high-level functional aspects so that our PET can be implemented as software running on the user's machine, for example, in the form of a Web browser add-on. Our assumptions about the proposed architecture are described next:

\begin{itemize}
  \item First, we assume that both the user and the adversary use the same time periods, for example, 24 uniformly distributed time slots within a day. This implies that the profile computed on the user's side coincides with the profile built by the attacker.

  \item Secondly, according to~\eqref{eq2}, our approach needs the user's actual profile $q$ to compute the optimal storing and forwarding strategies. Because of this, we contemplate a training period before our architecture starts delaying messages. However, since the attacker might learn about the user profile during this training period, the user could alternatively provide the software with an estimate of their profile.

  \item Lastly, we suppose that, in the estimation of the relative histogram, the components of the user profile remain stable after the training phase. We acknowledge, however, that a practical implementation of our mechanism should take into account that the user activity may vary significantly over time.
\end{itemize}

Before we proceed with the description of our architecture, we shall provide an example showing what the optimal storing and forwarding strategies mean in practice.
For this, consider the profile $q$ depicted in Fig.~\ref{fig3}(a), which corresponds to a user with initial privacy risk $\cP(0)\simeq 4.2775$ bits. If this user decided to delay $\varphi=4\%$ of their messages, the relative privacy gain would be around 5.18\%. That is, in this particular case we observe that the privacy gain would be, interestingly, greater than the delay rate introduced.

The optimal strategies are illustrated in Fig.~\ref{fig3}(b). The storing strategy suggests buffering 3.37\% and 0.63\% of messages at time instants 1 and 2, respectively\footnote{Those time instants are, in fact, time periods of one hour each. In particular, the time index $i$ consists in the interval $(i-1,i]$.}. On the other hand, the forwarding strategy recommends extracting 0.84\% of the total number of messages from the buffer at time periods 7, 8, 9 and 10, and 0.64\% of the messages at time 13.

\begin{figure}[t!]%
\centering
\includegraphics[scale=0.73]{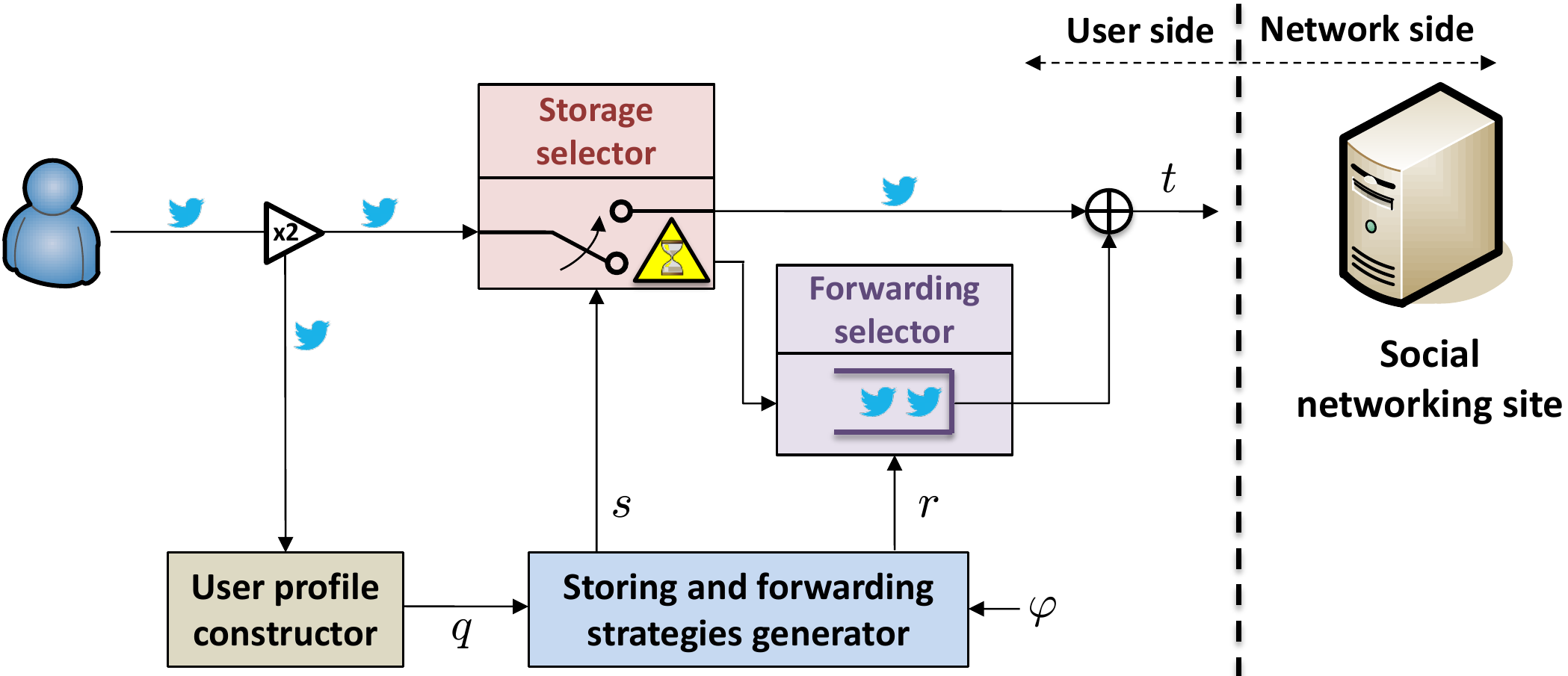}
\caption{Architecture implementing the message-deferral mechanism.}
\label{fig4}
\end{figure}

In Fig.~\ref{fig4} we depict the proposed architecture, which consists of a number of modules, each of them performing a specific task. From a general perspective, this figure shows a user interacting with a social networking site, an entity that basically stores the messages generated by this and other users. Next, we provide a functional description of the modules of this architecture.

\begin{ItemizeStandard}
  \item \textbf{User-profile constructor}. It is responsible for the estimation of the user's profile. Specifically, this module receives the messages the user generates, and computes a histogram of relative frequencies of these messages within, for example, 1-hour slot throughout one day. Afterwards, this profile is submitted to the \emph{storing and forwarding strategies generator}.

We would like to emphasize that this module is active even when the user explicitly declares their profile. Since the profile specified by the user may not be an accurate reflection of their online behavior, our architecture may decide, after the training phase, to replace it with the profile implicitly inferred from the posted messages.

  \item \textbf{Storing and forwarding strategies generator}. This module is the core of the architecture as it is responsible for computing the solution to the optimization problem inherent in function~\eqref{eq2}. To this end, this component is first provided with the user profile and the message-deferral rate. Secondly, this module computes the optimal tuples of storing and forwarding; and finally, those tuples are given to the \emph{storage selector} module and to the \emph{forwarding selector} block.

  \item \textbf{Storage selector}. The functionality of this module is to warn the user when they should delay messages\footnote{This would be, in fact, transparent to the user. The software installed on the user's machine would decide whether a message is to be delayed or not.}. Specifically, at time period $i$, with probability $s_i/q_i$ the user should send a message to the buffer implemented in the \emph{forwarding selector} module. On the other hand, with probability $1-s_i/q_i$, this message should be submitted directly to the social networking site.

  \item \textbf{Forwarding selector}. This block includes a buffer where messages are stored. Its main functionality is to output messages from this buffer according to the optimal forwarding strategy $r$. In particular, this module would operate as follows: throughout time slot $i$, the module would send $\alpha\,r_i$ messages from the buffer to the service provider, where $\alpha$ represents the total number of messages generated within the time period covered by the profile, e.g., one day.

This block also considers the possibility of assigning priorities to messages. For instance, it could be necessary that certain messages stored in the buffer have different levels of priority. As an example, those messages generated during working hours could have a higher likelihood of leaving the buffer. Other alternatives include first in, first out (FIFO), last in, first out (LIFO) and uniformly-random extraction. This last option is precisely the one considered in Sec.~\ref{sec:DelayCapacity}.
\end{ItemizeStandard}

\section{Expected Delay and Message-Storage Capacity}
\label{sec:DelayCapacity}
\noindent
In Sec.~\ref{sec:Technique:Formulation} we characterized the optimal privacy-utility trade-off posed by message deferral, in terms of the Shannon entropy of the apparent profile as measure of privacy, and the message-deferral rate as measure of utility. In that same subsection, we also mentioned that the optimization problem characterizing this trade-off is a particular case of a more general optimization problem for which there exists a closed-form solution. Although this allows us to obtain analytically our optimal storing and forwarding strategies for a given deferral rate, users would certainly benefit from more meaningful metrics of loss in usability than this fraction of messages delayed. In other words, it would be interesting and even necessary to investigate more elaborate and informative utility measures, capturing the actual impact that our mechanism would have.

Motivated by this, in this section we examine more sophisticated metrics such as the expected delay experienced by messages and the capacity of the buffer where these messages are stored. Further, we investigate how they relate each other, under the premise that messages are output from the buffer uniformly at random, that is, without considering any kind of priority such as FIFO or LIFO.

This section is structured as follows. First, Sec.~\ref{sec:DelayCapacity:Preliminaries} examines some interesting results derived from the more general optimization problem examined in~\cite[\S 7]{Parra13PhD}. Then, Sec.~\ref{sec:DelayCapacity:Theory} presents a mathematical analysis modeling the utility metrics mentioned above, namely, the expected delay and buffer capacity. Finally, Sec.~\ref{sec:DelayCapacity:Example} provides an example illustrating the theoretical results obtained in the previous subsection.

\subsection{Preliminaries}
\label{sec:DelayCapacity:Preliminaries}
\noindent
The optimization problem investigated in~\cite[\S 7]{Parra13PhD} is a resource allocation problem that arises in the context of privacy protection in recommendation systems. In the cited work, the authors model the privacy-utility trade-off posed by a data-perturbative mechanism consisting in the forgery and the elimination of ratings. Specifically, the privacy risk $\cR$ is measured as the KL divergence between the apparent profile of \emph{interests}\footnote{Here users' profiles do not capture their interests, but their online activity.} and the population's distribution of items $p$. On the other hand, the loss in accuracy of recommendations is measured as the percentages of ratings $\rho$ and $\sigma$ that the user would be willing to forge and suppress, respectively. Accordingly, the optimal trade-off between privacy and utility is defined as

\begin{equation}
\label{eq3}
\cR(\rho,\sigma)=  \min_{\substack{r,s\\r_i \geqslant 0, \, s_i \geqslant 0,\\ q_i  - s_i + r_i\geqslant 0,\\
\sum s_i = \sigma, \,\sum r_i = \rho}} \oD\left(\left. \frac{q-s+r}{1-\sigma+\rho}\,\right\|\,p\right),
\end{equation}
where the optimization variables are a \emph{forgery strategy} $r$ and a \emph{suppression strategy}~$s$.

In light of this formulation, it is straightforward to check, by virtue of~\eqref{eq1}, that
$$\cP(\varphi) = \log n - \cR(\varphi,\varphi)|_{p=u}.$$
In words, the function~\eqref{eq2} characterizing the trade-off between privacy and message-deferral rate is a special case of the optimization problem~\eqref{eq3}, when the rates of forgery and suppression are equal to~$\varphi$ and the population's distribution is the uniform distribution.
In the context of our formulation, the forgery and suppression strategies clearly correspond to the forwarding and storing strategies, respectively.

Having shown then that~\eqref{eq2} is a particular case of~\eqref{eq3}, next we review a couple of results presented in~\cite[\S 7]{Parra13PhD} to be used in the coming sections.

The most relevant result is the intuitive principle that the optimal storing and forwarding strategies follow.
Specifically, the former strategy lowers the highest values of $q_i$ until these values are equal. This is done in such a way that the values lowered amount to~$\varphi$. In a completely analogous manner, the latter strategy raises the lowest values of~$q_i$ until they match, for a total probability mass increment of~$\varphi$. Finally, intermediate values of~$q_i$ remain unperturbed. Simply put, the effect of the optimal strategies on the actual user profile may be regarded as a combination of the well-known water-filling and reverse water-filling problems~\cite[\S 5.5]{Boyd04B}.

\begin{figure}[t!]
\centering
\includegraphics[scale=\SinglePlotScale]{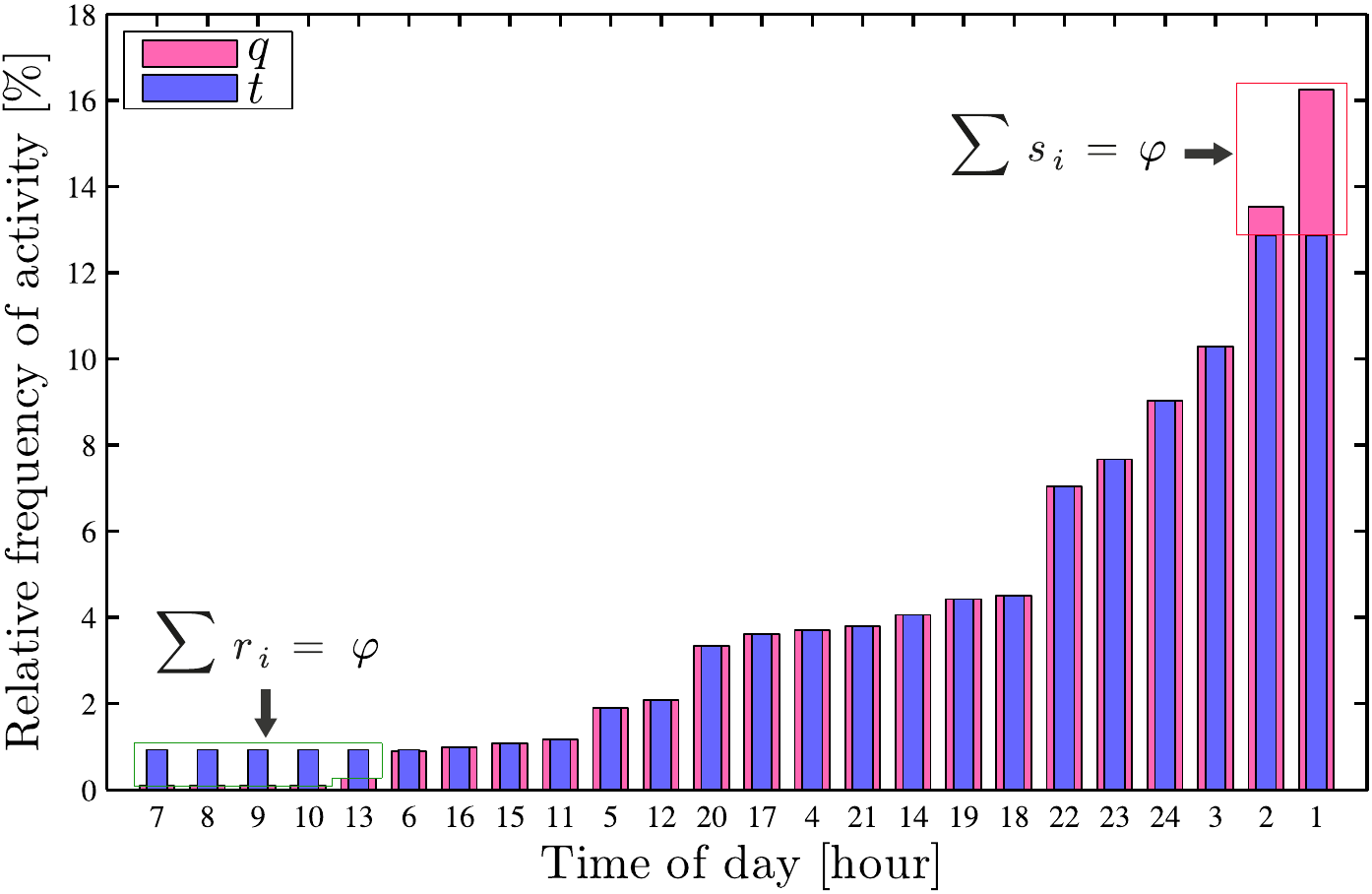}
\caption{This figure illustrates the intuition behind the optimal storing and forwarding strategies. Here, we have represented the actual user profile $q$ depicted in Fig.~\ref{fig3}(a). The optimal apparent profile $t$ is obtained by applying the strategies shown in Fig.~\ref{fig3}(b), which correspond to a deferral rate~$\varphi=0.04$.}
\label{fig5}
\end{figure}

The aforementioned principle was already anticipated in Fig.~\ref{fig3}. In Fig.~\ref{fig5} we illustrate this more clearly. Particularly, this figure depicts the actual user profile shown in Fig.\ref{fig3}(a) and its optimal apparent profile, resulting from the application of the optimal storing and forwarding strategies represented in Fig.~\ref{fig3}(b). In Fig.~\ref{fig5}, however, the components of those two profiles are sorted in increasing order of activity to emphasize the way these strategies operate.

Another interesting result from~\cite[\S 7]{Parra13PhD} confirms the intuition that there must exist a pair~$(\rho,\sigma)$ such that the privacy risk vanishes. In the context of our formulation, this implies that there is a deferral rate $\varphi$ beyond which the maximum level of privacy or \emph{critical privacy} is attained\footnote{Recall from
Sec.~\ref{sec:Technique:Metric} that Shannon's entropy is regarded here as a measure of privacy \emph{gain}, whereas the KL divergence is interpreted as a measure of privacy \emph{risk}.}. We refer to this rate as the \emph{critical message-deferral rate} $\phicrit$.

Recall~\cite{Cover06B} that the \emph{variational distance} between two PMFs $p$ and $q$ is defined as
$$\textnormal{TV}(p\,\|\,q) = \tfrac{1}{2} \sum_i  \left|p_i - q_i\right|.$$
It can be shown~\cite[\S 7]{Parra13PhD} that the critical rate yields
$$\phicrit = \textnormal{TV}(u\,\|\,q).$$
From this expression, it is easy to verify that $\phicrit \geqslant 0$, with equality if, and only if, $q=u$. Later on, in Sec.~\ref{sec:Experiments}, we shall determine the average critical rate within a population of Twitter users, as well as the PMF of this crucial parameter.

The last result is related to the orthogonality of the components of $s$ and $r$. Specifically, it follows from~\cite[\S 7]{Parra13PhD} that, for any $\varphi \leqslant \phicrit$, the optimal storing and forwarding strategies satisfy
$$s_k\,r_k=0,$$
for $k=1,\ldots,n$. The orthogonality of both strategies, in the sense indicated above, conforms to intuition --- it would not make any sense to store messages in a given time period and, at the same time period, forward messages to the social networking server. This result is implicitly assumed throughout next subsection, Sec.~\ref{sec:DelayCapacity:Theory}.

\begin{figure}[t]%
\centering
\includegraphics[scale=\SinglePlotScale]{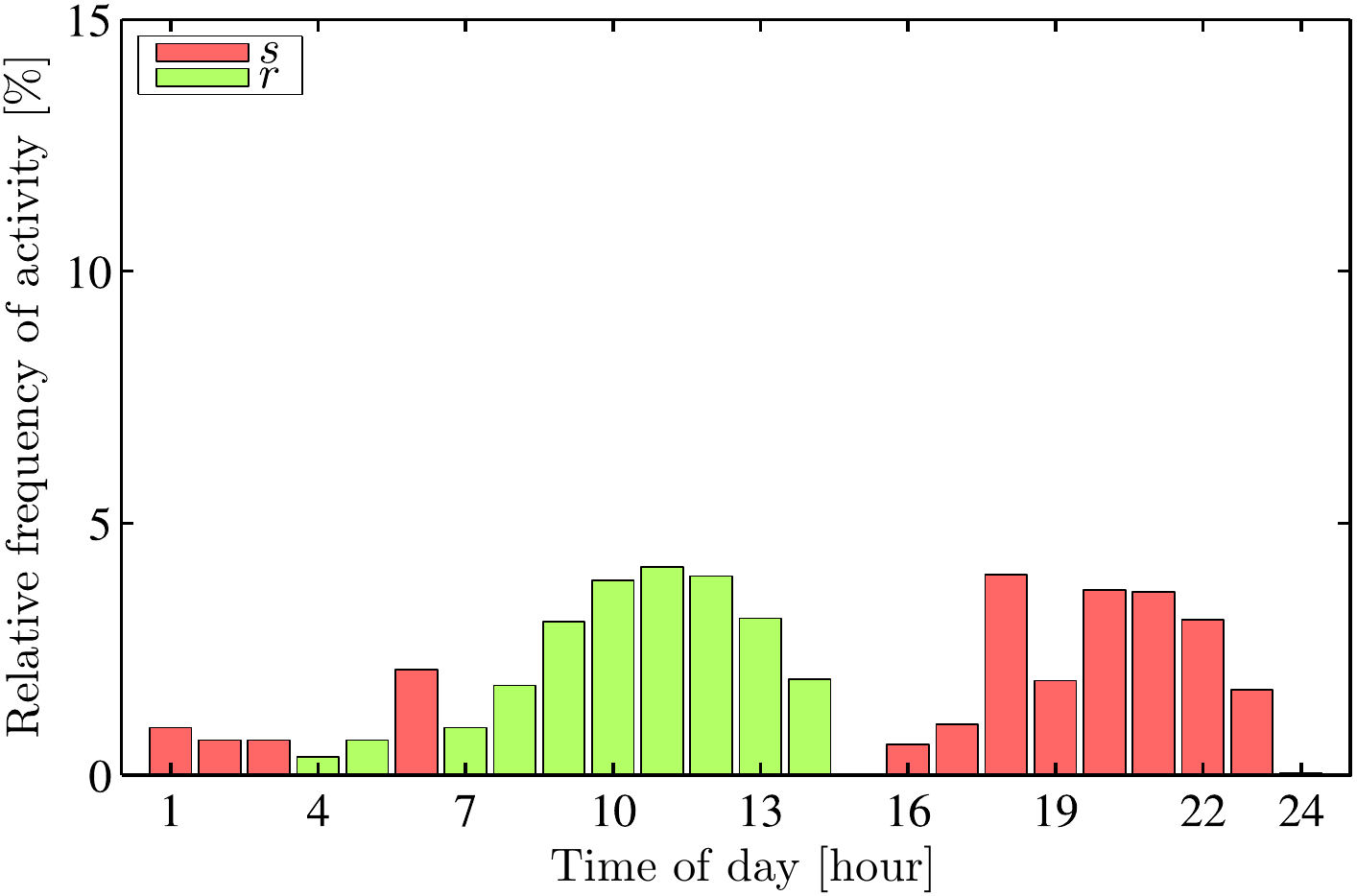}
\caption{Example of optimal storing and forwarding strategies which do not satisfy the principle of causality.}
\label{fig6}
\end{figure}

\subsection{Theoretical Analysis}
\label{sec:DelayCapacity:Theory}
\noindent
Denote by $s=(s_1,\ldots,s_n)$ and $r=(r_1,\ldots,r_n)$ the solutions to the problem~\eqref{eq2}, conceptually, a storing strategy and a forwarding strategy that maximize the Shannon entropy of the apparent profile, $\oH(t)$. Recall that these two tuples must satisfy
\begin{equation}\label{eq4}
\sum s_i = \sum r_i = \varphi.
\end{equation}

In Fig.~\ref{fig3}(b) we depicted an example of these tuples. In that figure, the time instants when messages were stored were preceding the time instants when these messages were forwarded. That is, the figure showed the logical sequence in which messages are first kept in the buffer and then they are flushed out.

However, the solutions $s$ and $r$ do not need to satisfy this principle of causality; this was not specified as a constraint in the optimization problem~\eqref{eq2}. In fact, regardless of whether causality is satisfied or not, these two tuples must be interpreted as cyclic sequences, which are repeated continuously, e.g., every day or week, depending on the time frame covered by the user profile. This is how the storing and the forwarding strategies must be construed then in Fig.~\ref{fig6}. Here, although no messages are forwarded at the time instants 1, 2 and 3 of the first cycle (day), in subsequent cycles these time instants will be used to output messages.

In the remainder of this section, we shall mathematically model the buffer. Specifically, we shall find a time instant such that, if the tuples $s,\,r$ are moved to start at this instant, then every message to be forwarded during the next $n$ consecutive time periods will actually be forwarded. This is time period 7 in the example shown in Fig.~\ref{fig6}. With this time index, we shall be able to proceed to find an expression for the expected delay.

Denote by $a_i$ the $n$ consecutive permutations of the tuple $s_i-r_i$,
\begin{equation*}
\setlength\arraycolsep{0.05em}
 \begin{array}{lcccccccr}
\vspace{0.1cm}
a_1=(&s_1-r_1&,& s_2-r_2&, &\ldots&,&s_n-r_n&),\\
\vspace{0.1cm}
a_2=(&s_2-r_2&,&\ldots&, &s_n-r_n&,&s_1-r_1&),\\
\,\,\,\vdots\\
\vspace{0.1cm}
a_n=(&s_n-r_n&,&s_1-r_1&,&\ldots&,&s_{n-1}-r_{n-1}&).
\end{array}
\end{equation*}
Associated with each tuple $a_i$, define the sequence~$\left(b_{i,j}\right)_{j=1}^{\infty}$ as
$$b_{i,j}=\left\{ \begin{array}{r@{,\quad}l}
                                            \max\{a_{i,j},0\} & j=1 \\
                                            \max\{b_{i,j-1}+a_{i,k},0\} & j= 2,3,\ldots
\end{array}\right.,$$
where $k=j-n \lfloor\frac{j-1}{n}\rfloor$ indexes cyclically the tuple~$a_i$.
Conceptually, each sequence $b_i$ models the ratio of messages to total number of messages that are stored in the buffer over the time index $j$, when the optimal storing and forwarding strategies are applied cyclically starting from the time index~$i$.

Recall that the Heaviside step function~\cite{Apostol74B} of a discrete variable $x$ is defined as
$$\theta(x)=\left\{ \begin{array}{l@{,\quad}l}
                                            0 & x < 0 \\
                                            1 & x\geqslant 0
\end{array}\right..$$
Our first result demonstrates that, after a transient state and regardless of the starting index $i$, these sequences converge to a common, repeated pattern. As we show next, this is a consequence of~\eqref{eq4}.
\SpaceAfterPropositionEndedWithFormula
\begin{lemma}\label{lemma}
\leavevmode
\begin{EnumerateRoman}
\item There exists some index $i=1,\ldots,n$ such that $b_{i,n}=0$.
\item Let $i$ be an index satisfying $b_{i+1,n}=0$. Then, for any $j=1,\ldots,n$, there exists an index~$l=i+1-j+n\,\theta(j-i-1)$ such that
$$b_{j,k+l} = b_{i+1,k}, \textnormal{   for } k=1,2,\ldots$$
\end{EnumerateRoman}
\end{lemma}

\BeginProof
Define the $n$ cumulative sums $w_{i,j}=\sum_{k=1}^j a_{i+1,k}$  for $i=1,\ldots,n-1$, and $w_{i,j}=\sum_{k=1}^j a_{i-n+1,k}$  for $i=n$, where the index $j$ ranges from 1 to $n$. Note that $w_{i,j}=w_{n,i+j}-w_{n,i}$, for $i+j\leqslant n$; when $i+j>n$, substitute $i+j$ for $i+j-n$. Let $i\in\{1,\ldots,n\}$ be an index such that $w_{n,i}$ is minimal. Then, it immediately follows that $w_{i,j} \geqslant 0$ for $j=1,\ldots,n$, which implies, by virtue of~\eqref{eq4}, that $b_{i+1,n}=0$. Note that this holds also for the index $i=n$, for which $b_{1,n}=0$. This proves statement~(i).

We have showed that the index $i$ that minimizes $w_{n,k}$ for all $k$ satisfies $b_{i+1,n}=0$. To prove~(ii), first we shall show that $b_{j,l}=0$ for all $j$. To this end, replace the index $j$ with $j+1$ in statement~(ii), so that now $l=i-j+n\,\theta(j-i)$ and $j$ goes from 0 to $n-1$. Recall also that $w_{i,j}=w_{n,i+j}-w_{n,i}$ for $i+j \leqslant n$, and $w_{i,j}=w_{n,i+j-n}-w_{n,i}$ for $i+j>n$. With the previous change of variable, note that $w_{j,l}=w_{n,i}-w_{n,j}$. Here, for consistency with the indexes of $w_{i,j}$, we substitute $j=0$ for $j=n$. Having said this, observe that, for a given~$j$
$$\min_k w_{j,k}= w_{n,i} - w_{n,j}=w_{j,l},$$
which clearly is nonpositive. Then, fix $j$ and note that the set of possible values that $b_{j+1,k}$ may take on are $w_{j,k}$ plus the $k$ terms $w_{j,k}-w_{j,m}$ for $m=1,\ldots,k$. Since $\min_k w_{j,k}=w_{j,l} \leqslant 0$, it follows that $b_{j+1,l}=0$. To conclude the proof, simply observe that $a_{j+1,l+1}=a_{i+1,1}$.~\EndProof
\SpaceAfterPropositionEndedWithFormula

Hereafter we shall refer to the \emph{starting index} $i$ as the index satisfying $b_{i,n}=0$.
Since Lemma~\ref{lemma} shows that all sequences converge to a steady state where a pattern is repeated continuously, our analysis is restricted to the finite sequence $\left(b_{i,j}\right)_{j=1}^n$ modeling this pattern and its corresponding tuple $a_i$.

Let $C$ be the capacity of the buffer and $\alpha$ the total number of messages generated by the user throughout the considered time frame (a day, week, month, etc.). The next result gives a straightforward expression for $C$ when the steady state is achieved.
\SpaceAfterPropositionEndedWithFormula
\begin{corollary}\label{corollary}
\leavevmode
Let $i$ be the starting index. In the steady state, the buffer capacity is
$$C = \alpha \max_{j\in \{1,\ldots,n\}}  b_{i,j}.$$
\end{corollary}
\BeginProof
It is immediate from the definition of $b_{i,j}$ and Lemma~\ref{lemma}.
\EndProof
\SpaceAfterPropositionEndedWithFormula

Next, we shall reorder the tuples $s,r$ so that they begin at the starting index. Denote by $s'$,$r'$ the tuples starting with this index $i$, formally
$$s'=(s'_1,s'_2,\ldots,s'_n),$$
$$r'=(r'_1,r'_2,\ldots,r'_n),$$
where $s'_k=s_l$ and $r'_k=r_l$ with $l=i+k-1-n\lfloor \frac{i+k-2}{n}\rfloor$. Note that, when we reorder the storing and forwarding tuples this way, for every $r'_j>0$ we can forward exactly $r'_j$ messages at time period~$j$.

In the following we define some notation that will be used in Theorem~\ref{theorem}. Let $\Delta$ be an r.v.\ representing the number of time periods a message is delayed. Note that, on account of Lemma~\ref{lemma}, the buffer does not retain any message for more than $n$ time units. Consequently, the alphabet of $\Delta$ is the set $\{1,\ldots,n\}$. Denote by $\bar{\delta}$ its expected value, $\oE\Delta$. Let $D$ be a Bernoulli r.v.\ of parameter $\varphi$, modeling whether a message is delayed or not. Namely, $\oP\{D=1\}=\varphi$ is the probability that a message is delayed and $\oP\{D=0\}=1-\varphi$ is the probability it is not. Finally, define
$$\omega(j,k)=\{l: r'_l>0,k<l<j\}.$$

Our next result, Theorem~\ref{theorem}, provides a closed-form expression to calculate the expected delay in the steady state.
\SpaceAfterPropositionEndedWithFormula
\begin{theorem}\label{theorem}
\leavevmode
Let $i$ be an index satisfying  $b_{i,n}=0$. Then,
\begin{equation*}
\bar{\delta} = \sum_{\delta=1}^n \delta \sum_{\substack{j-k=\delta,\\r'_j,s'_j>0}} \frac{r'_j\,s'_k}{b_{i,j-1}} \prod_{l \in \omega(j,k)} \left(1- \frac{r'_l}{b_{i,l-1}}\right).
\end{equation*}
\end{theorem}

\BeginProof
From Lemma~\ref{lemma}, we know that all sequences $b_k$ for $k=1,\ldots,n$ converge to the finite sequence $\left(b_{i,j}\right)_{j=1}^n$. Note that $\oE\Delta =\oE \oE [\Delta|D]=\varphi \oE_{\Delta|D} [\Delta | D=1]$.
Next, we proceed to calculate the conditional PMF $p_{\Delta|D}(\delta|1)$. Let $A$ be an r.v.\ representing the time instant when a message arrives at the buffer, and $L$, the time instant when this message leaves the buffer. Accordingly,
\begin{equation*}
\oP\{\Delta=\delta|D=1\}=\sum_{j-k=\delta} \oP\{L=j |A=k\} \oP\{A=k|D=1\}.
\end{equation*}

Observe that $\oP\{A=k|D=1\}=s'_k/\varphi$. Further, note that $\oP\{L=j |A=k\}$ is the probability that a message is not forwarded at the time instants $\omega(j,k)$, that is, $\prod_{l\in \omega(j,k)} \left(1-\frac{r'_l}{b_{i,l-1}}\right)$, multiplied by the probability that this message is forwarded at time $j$, that is, $\frac{r'_j}{b_{i,j-1}}$. From this, it is immediate to derive the expression given in the statement of the theorem.
\EndProof
\SpaceAfterPropositionEndedWithFormula

The expression obtained in Theorem~\ref{theorem} allows us therefore to estimate the expected delay that messages will experience for a given deferral rate. Although at first sight it may seem there is not a direct dependence on the parameter $\varphi$, recall that $s$ and $r$ are related to this parameter through~\eqref{eq4}.

In conclusion, the results provided in this subsection enable us to establish a connection between the message-deferral rate, i.e., our simplified, but mathematically tractable measure of utility, and more elaborate and informative utility metrics such as the expected delay and the message-storage capacity.

\subsection{Numerical Example}
\label{sec:DelayCapacity:Example}
\noindent
This subsection presents a numerical example that illustrates the analysis conducted in the previous subsection, and shows the privacy level achieved by a user who adheres to the proposed message-deferral mechanism. Throughout this subsection, all results correspond to the same user.

\begin{figure*}[b!]%
\centering\hspace*{\fill}
\subfigure[$\varphi=0$, $\oH(t)\simeq 4.1060$ bits.]%
{\includegraphics[scale=\FigScaleMultipleFour]{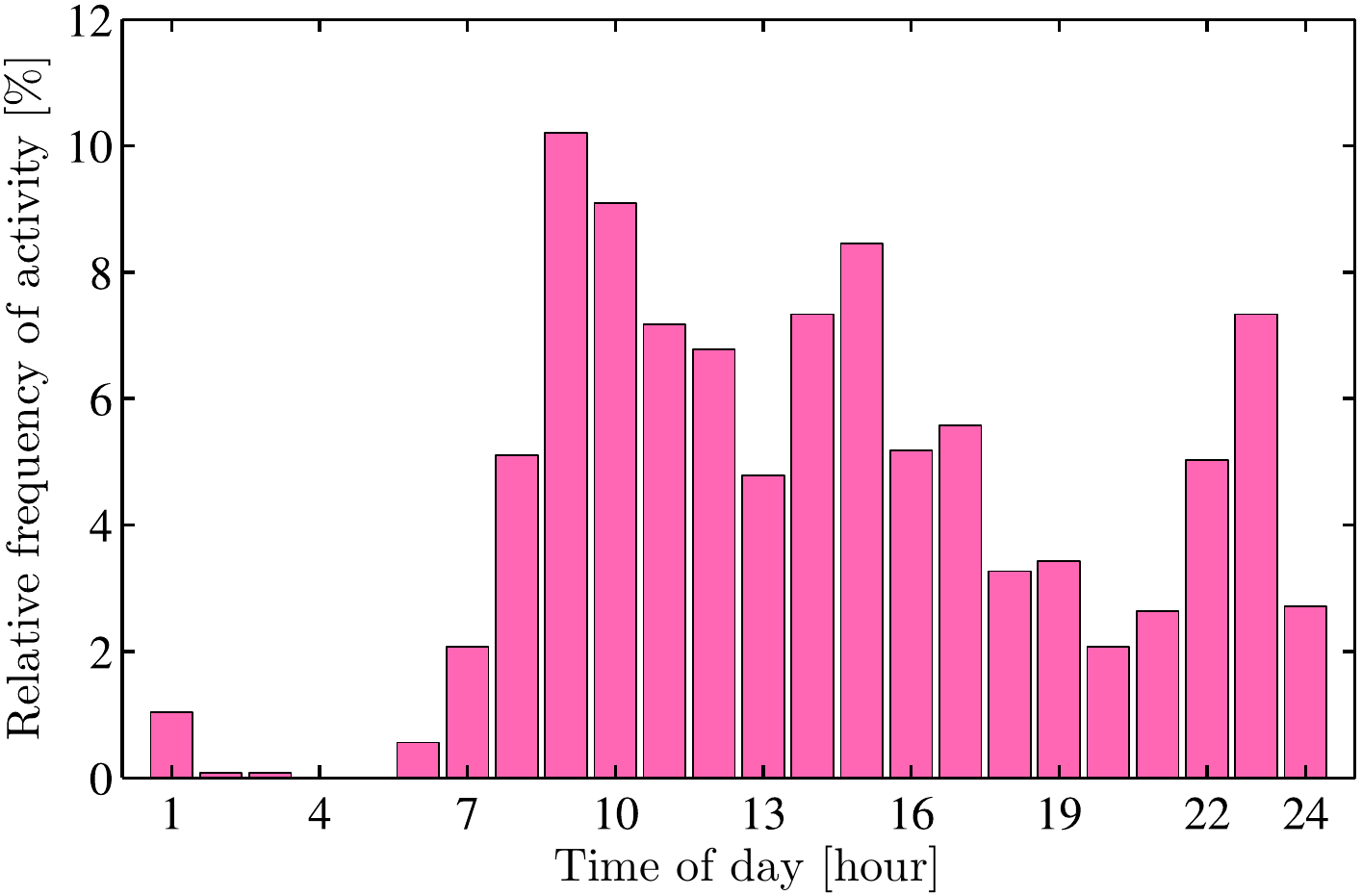}}%
\hfill
\subfigure[$\varphi\simeq0.1111$, $\oH(t)\simeq 4.4402$ bits.]%
{\includegraphics[scale=\FigScaleMultipleFour]{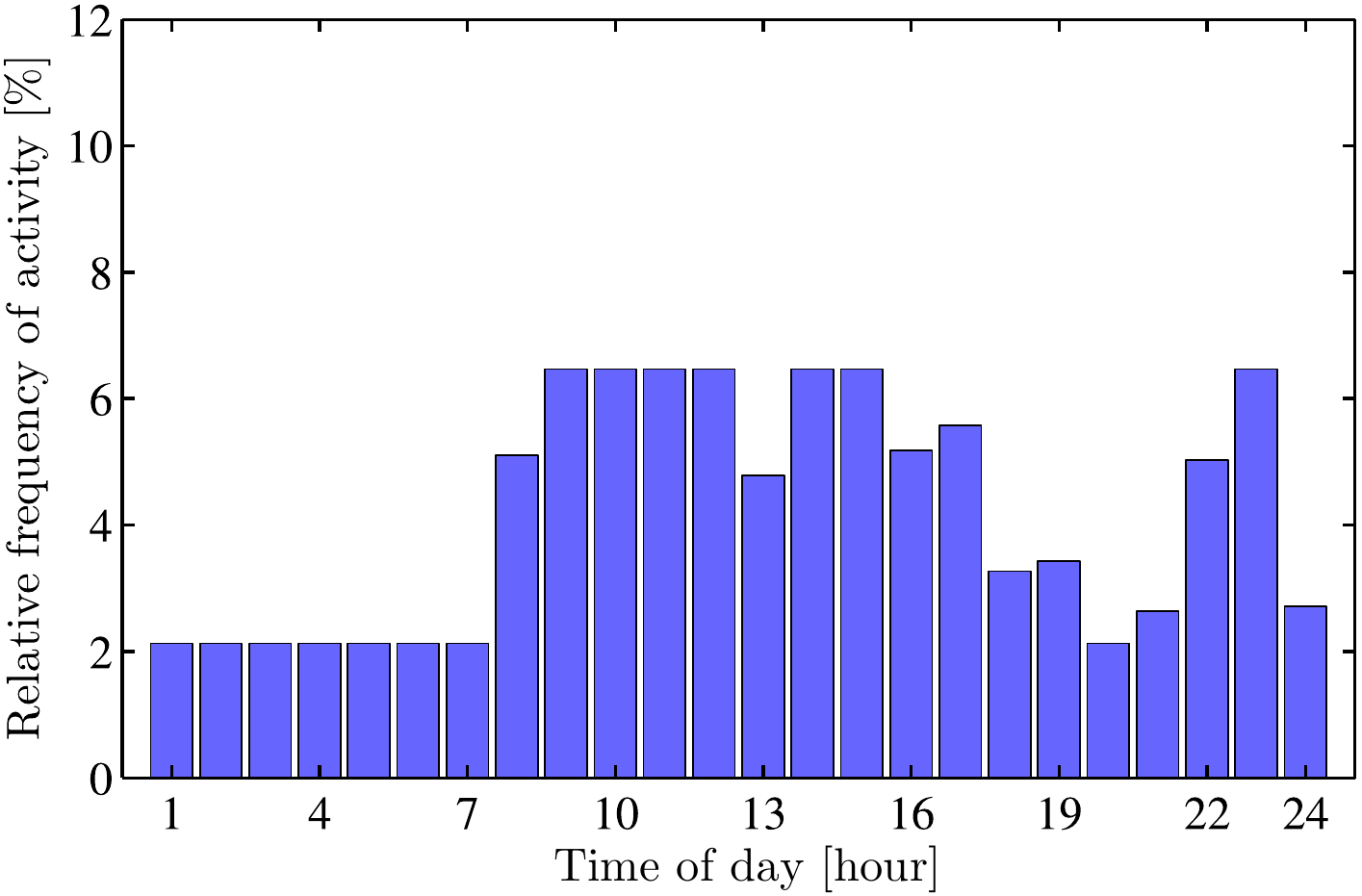}}%
\hspace*{\fill}
\\
\hspace*{\fill}
\subfigure[$\varphi\simeq0.2222$, $\oH(t)\simeq 4.5567$ bits.]%
{\includegraphics[scale=\FigScaleMultipleFour]{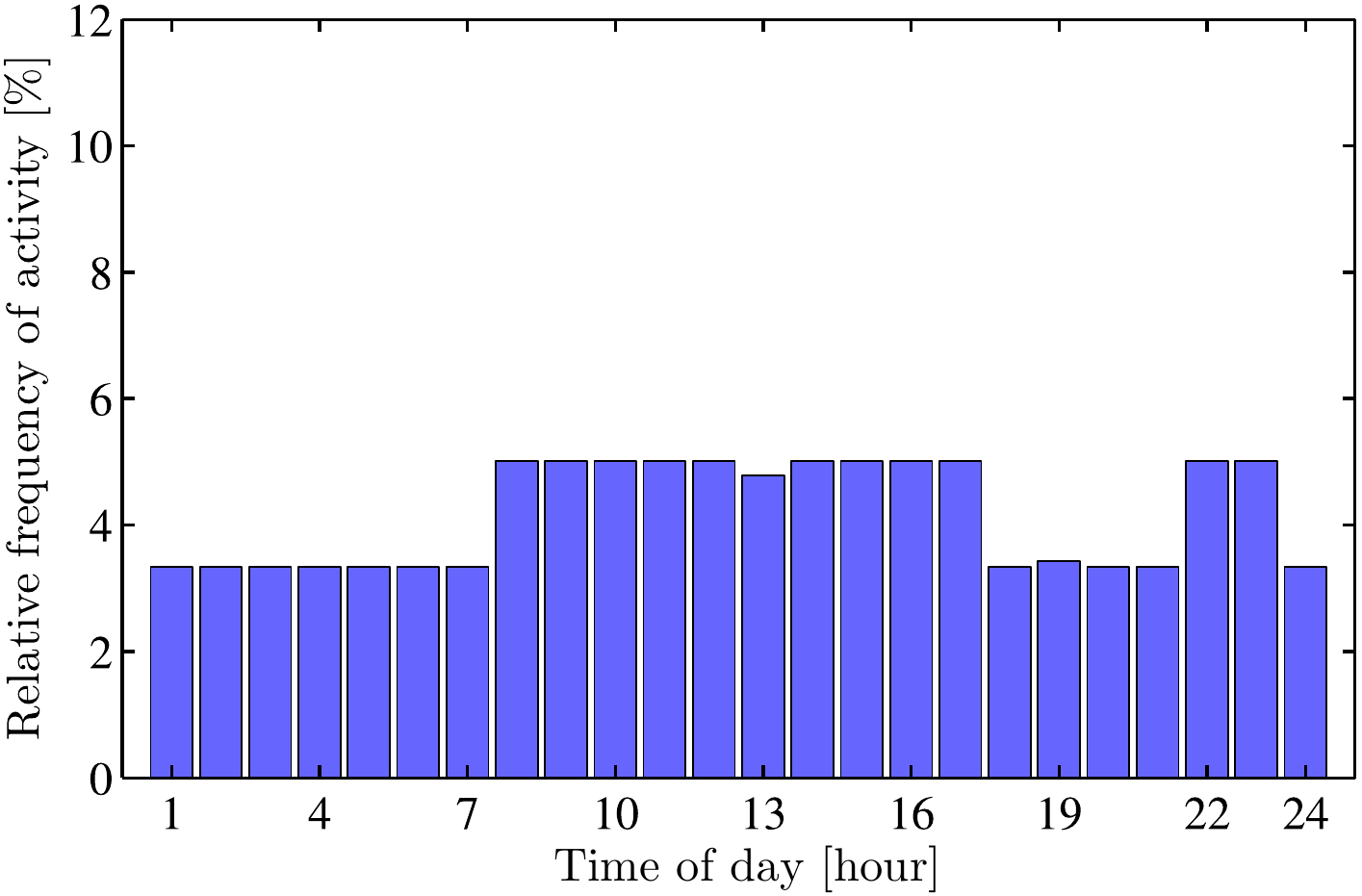}}%
\hfill
\subfigure[$\varphi\simeq0.3206$, $\oH(t)\simeq 4.5850$ bits.]%
{\includegraphics[scale=\FigScaleMultipleFour]{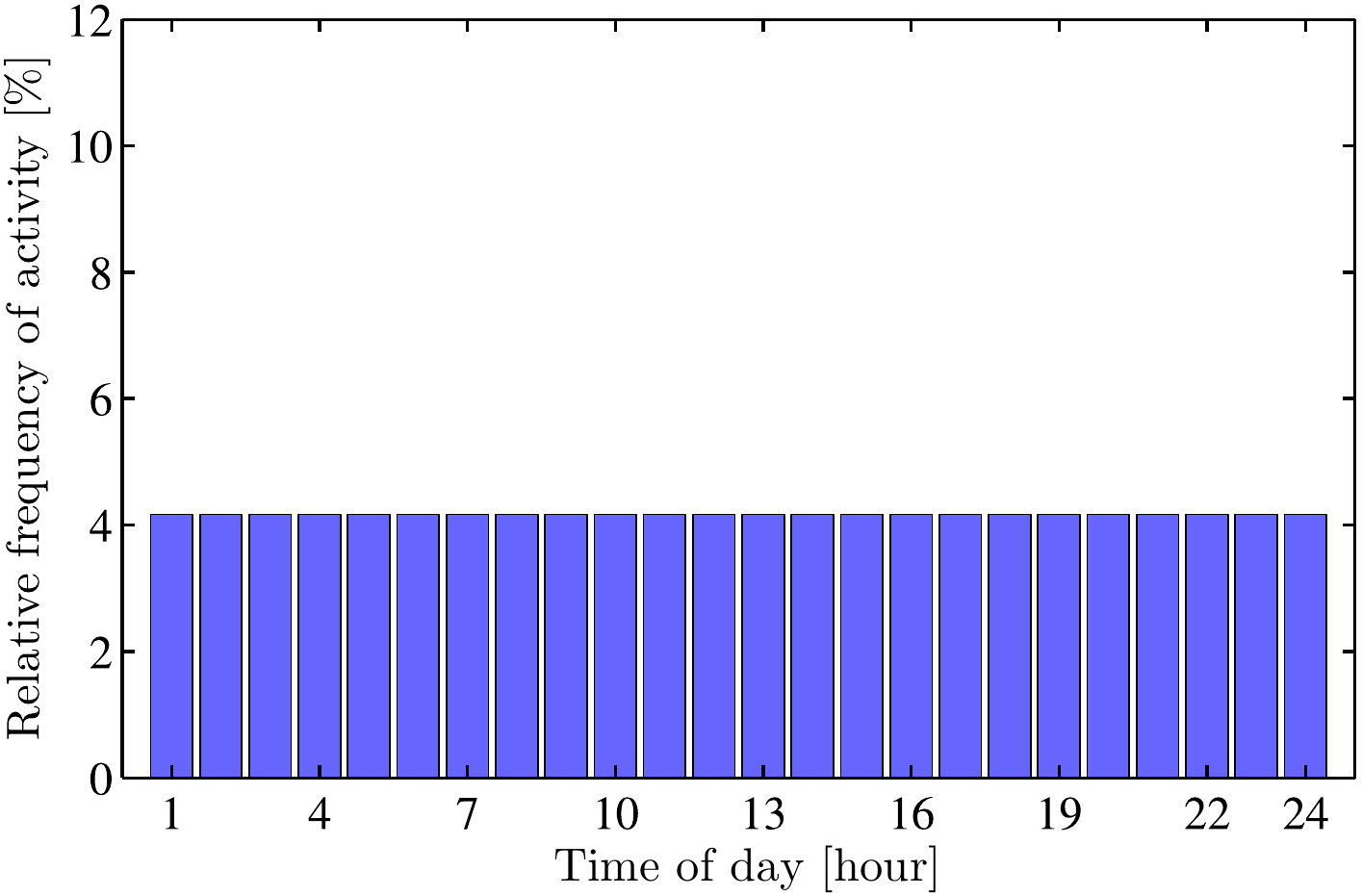}}%
\hspace*{\fill}
\caption{Apparent profiles for different values of $\varphi$.}
\label{fig7}
\end{figure*}

\begin{figure}[tb]%
\centering
\includegraphics[scale=\SinglePlotScale]{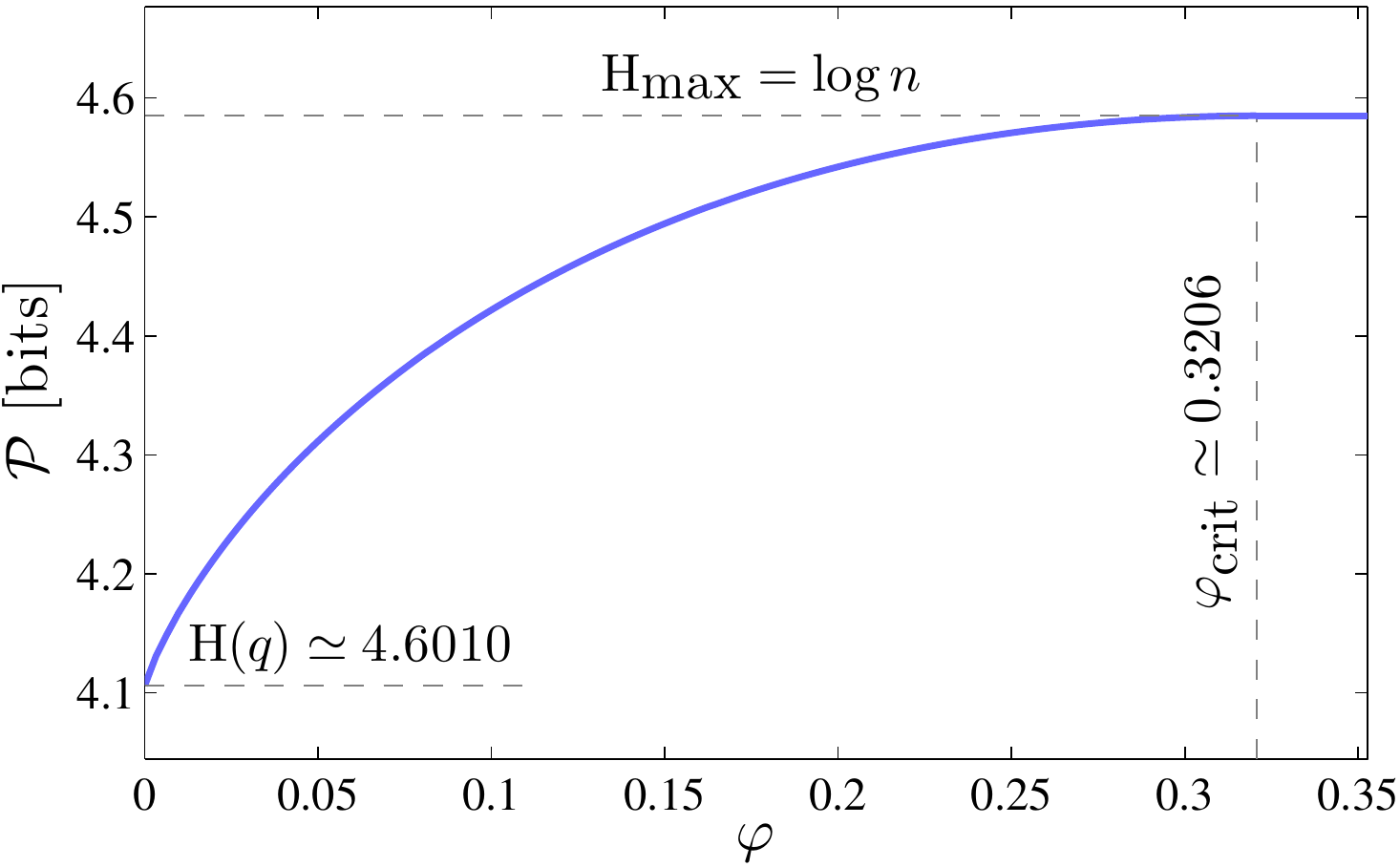}
\caption{Optimal trade-off between privacy and utility, the latter being measured as the message-deferral rate.}
\label{fig8}
\end{figure}

In Fig.~\ref{fig7} we represent the apparent profile of this user for different values of the message-deferral rate~$\varphi$. When $\varphi=0$, no perturbation takes place and the apparent profile $t$ represented in Fig.~\ref{fig7}(a) actually corresponds to the genuine user profile $q$. According to the reasoning behind the optimal storing and forwarding strategies described in Sec.~\ref{sec:DelayCapacity:Preliminaries}, the higher~$\varphi$, the more uniform is the resulting apparent profile. The maximum level of privacy is attained precisely for $\varphi=\phicrit \simeq 0.3206$, when the apparent profile is completely uniform and therefore $\oH(t)=\log24 \simeq 4.5850$. All this information is also captured in Fig.~\ref{fig8}, where we plot the privacy-deferral function~\eqref{eq2}, that is, the function modeling the optimal trade-off between privacy and utility, the latter being measured as the percentage of messages delayed.

 \begin{figure*}[tb]%
\centering\hspace*{\fill}
\subfigure[]%
{\includegraphics[scale=\FigScaleMultipleTwo]{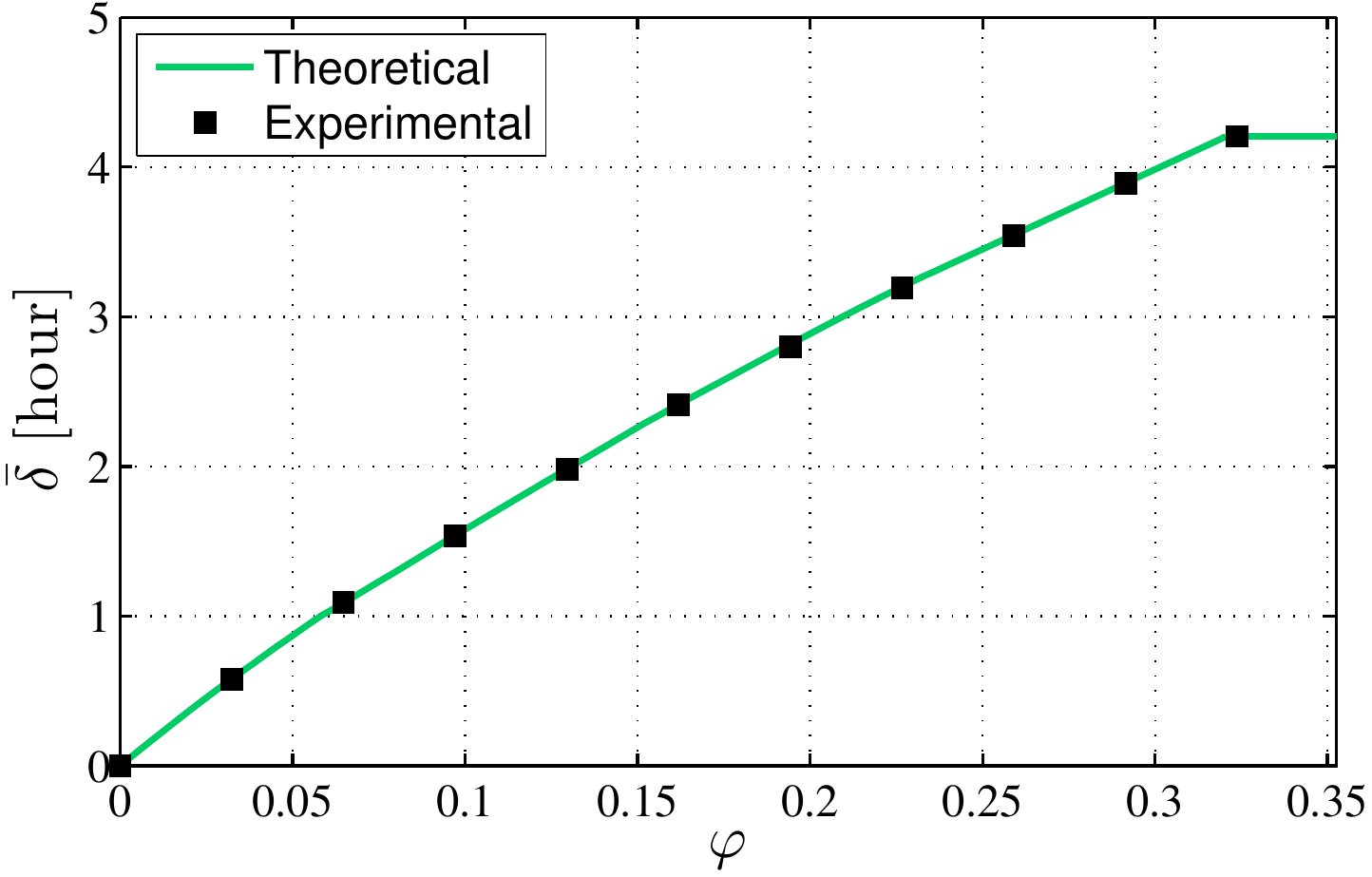}%
}\hfill
\subfigure[]%
{\includegraphics[scale=\FigScaleMultipleTwo]{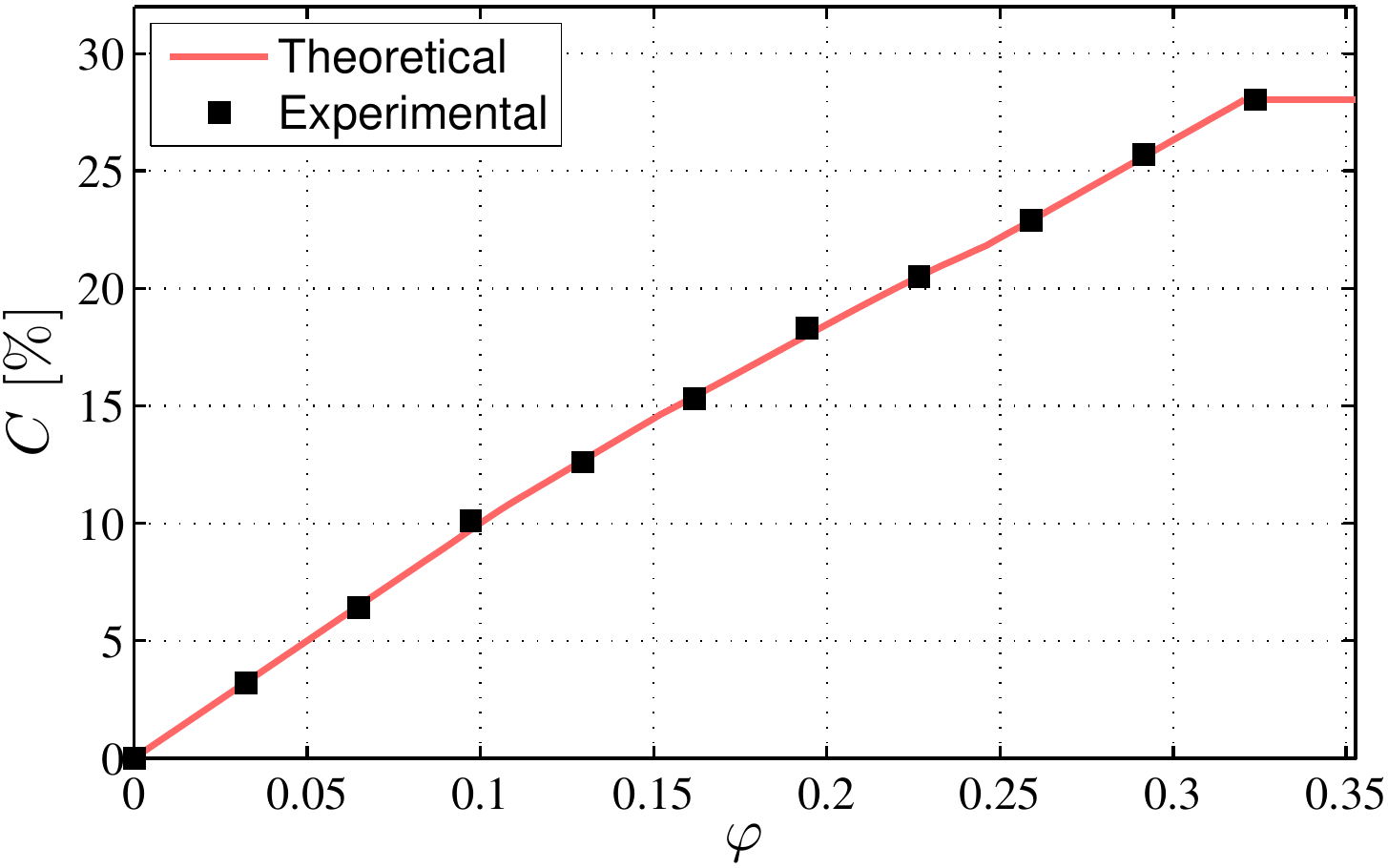}%
}\hspace*{\fill}
\caption{Expected delay (a) and buffer capacity (b) for different values of the message-deferral rate. The buffer requirements are expressed in relative terms, compared to the user's activity.}
\label{fig9}
\end{figure*}

In Fig.~\ref{fig9}(a) we depicted the expected delay $\bar{\delta}$ for different values of $\varphi$. In particular, the results shown in this figure were computed theoretically, by applying Theorem~\ref{theorem}, and experimentally. These latter experimental results were obtained by simulating the storing and forwarding processes as specified by the blocks \emph{storage selector} and \emph{forwarding selector} of the proposed architecture (see Sec.~\ref{sec:Architecture}). Fig.~\ref{fig9}(a) tells us, for example, that for $\varphi=0.10$, the messages delayed were kept on the buffer for around 1.5 hours on average. As expected, for $\varphi \leqslant \phicrit$, we observe that $\bar{\delta}$ exhibits an increasing, nonlinear behavior with $\varphi$. The case when $\varphi \geqslant \phicrit$ is of no interest as, in practice, a user would not delay more messages than those strictly necessary to achieve the maximum level of privacy.

Finally, Fig.~\ref{fig9}(b) shows, for different values of $\varphi$, the ratio between the number of messages stored in the buffer and the total number of messages generated by the user. For instance, when the user specifies $\varphi=0.10$, the buffer must be designed to keep around 10\% of all messages sent over a day. Clearly, we note that the buffer capacity is nonlinear with the deferral rate. Also, we observe that the user would need to store 28.1\% of their messages for the apparent profile to become the uniform distribution.

\section{Experimental Analysis}
\label{sec:Experiments}
\noindent
In this section we evaluate the extent to which the deferral of messages could enhance user privacy in a real-world scenario. The social network chosen to conduct this evaluation is Twitter, an online social networking platform that allows users to post messages of up to 140 characters.

In our experiments we employed 144 users, whose profiles were retrieved by using the Twitter API\footnote{\url{https://dev.twitter.com}}. In particular, we gathered the timestamps of all messages generated by those users before Oct.\ 25, 2013. From this information, we built their profiles as normalized histograms of tweets across 24 uniformly distributed time slots within one day. On average, users posted 1\,879.42 messages each.

\begin{figure}[b]%
\centering
\includegraphics[scale=\SinglePlotScale]{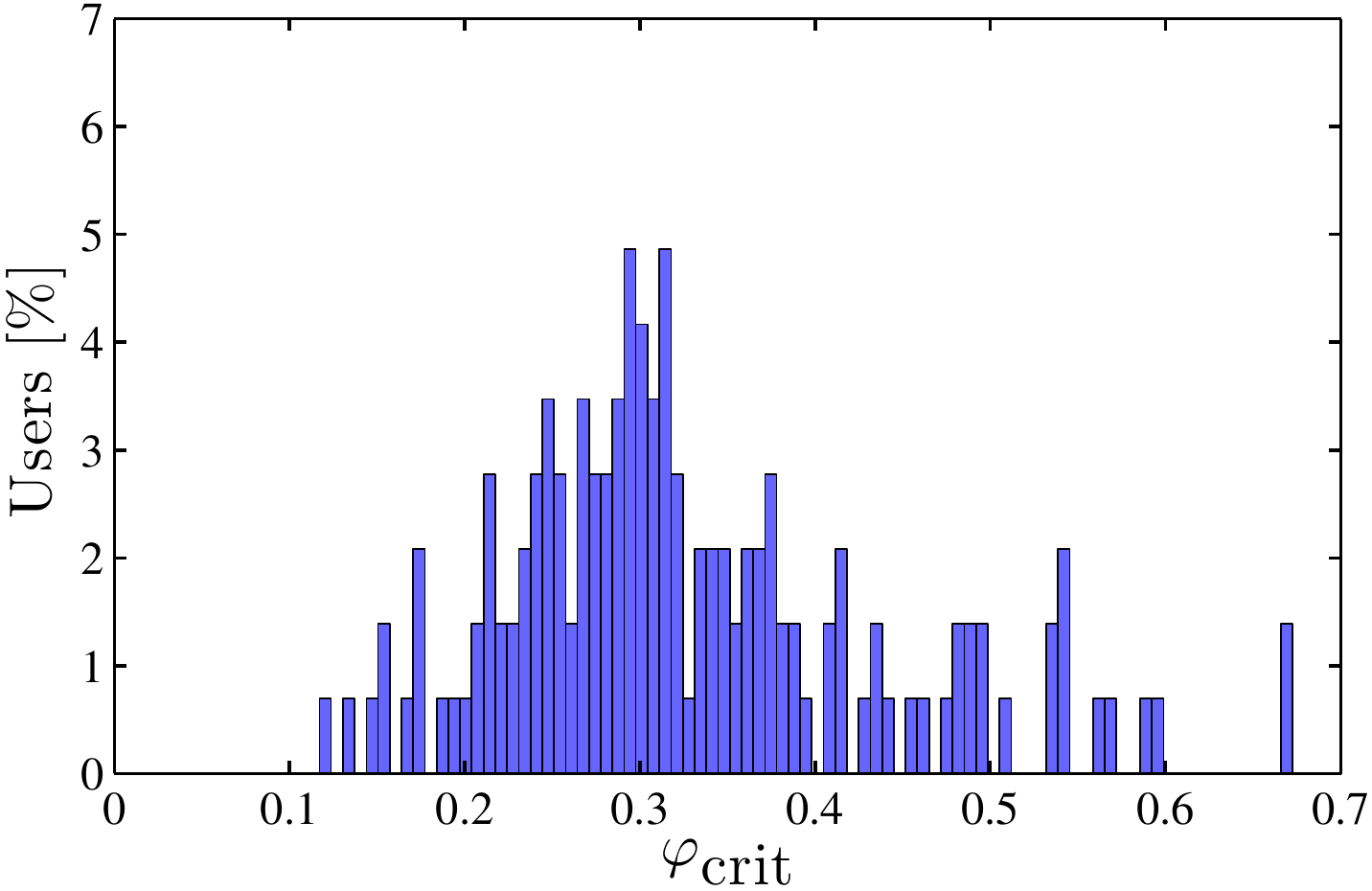}
\caption{Probability distribution of the critical message-deferral rate.}
\label{fig10}
\end{figure}

In our first series of experiments, we computed the probability distribution of $\phicrit$, that is, the message-deferral rate beyond which the maximum privacy level is achieved. The PMF of this critical rate is shown in Fig.~\ref{fig10}. As we can observe, the minimum and maximum values attained by $\phicrit$ are approximately 0.12 and 0.67. Also, we spot that a significant mass of probability is concentrated between $\varphi\simeq0.2$ and $\varphi\simeq0.4$, in particular, a 74\% of users. This means that most users will not require delaying a large percentage of their tweets for their apparent profiles to become the uniform distribution.

\begin{figure*}[tb]%
\centering\hspace*{\fill}
\subfigure[]%
{\includegraphics[scale=\FigScaleMultipleTwo]{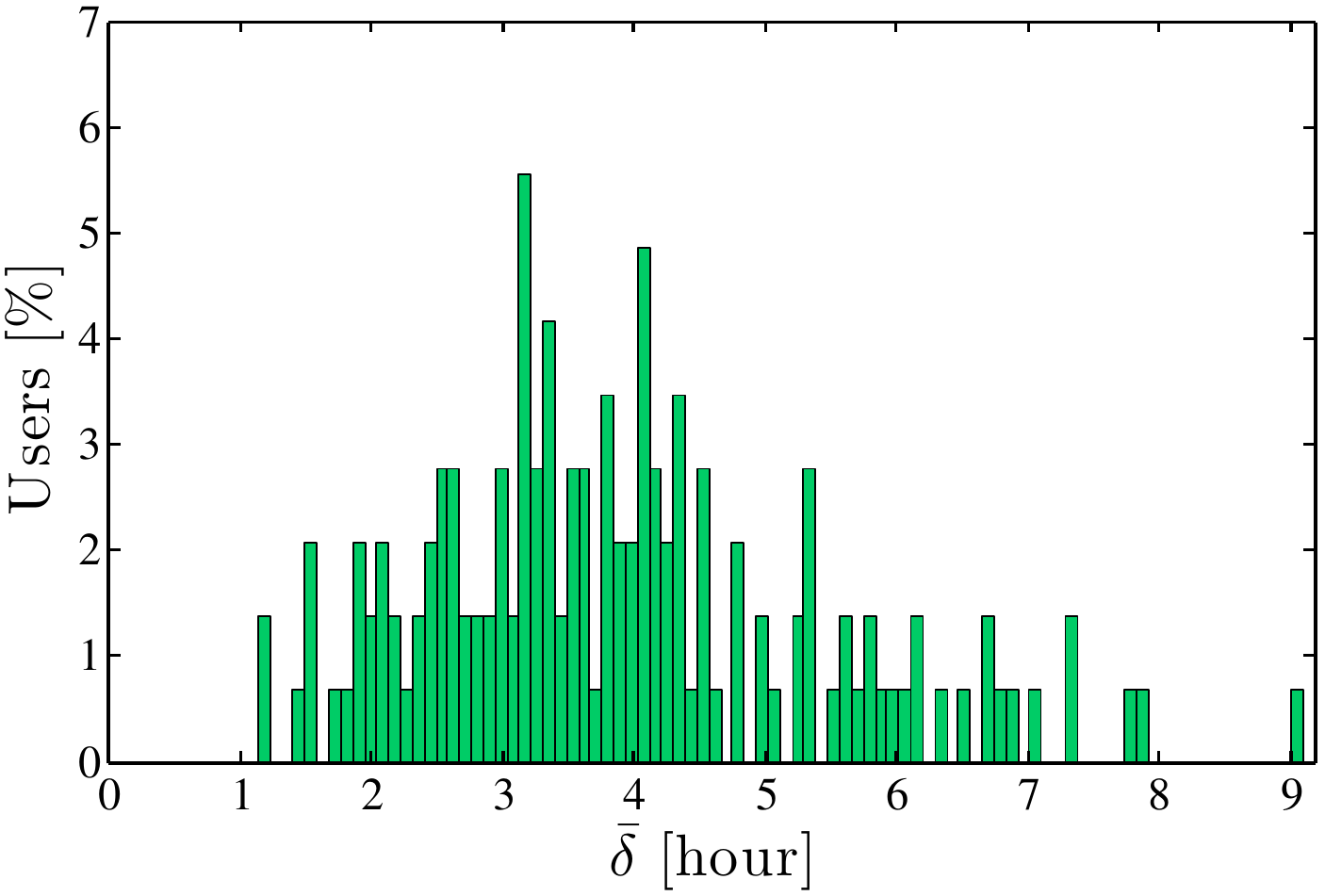}%
}\hfill
\subfigure[]%
{\includegraphics[scale=\FigScaleMultipleTwo]{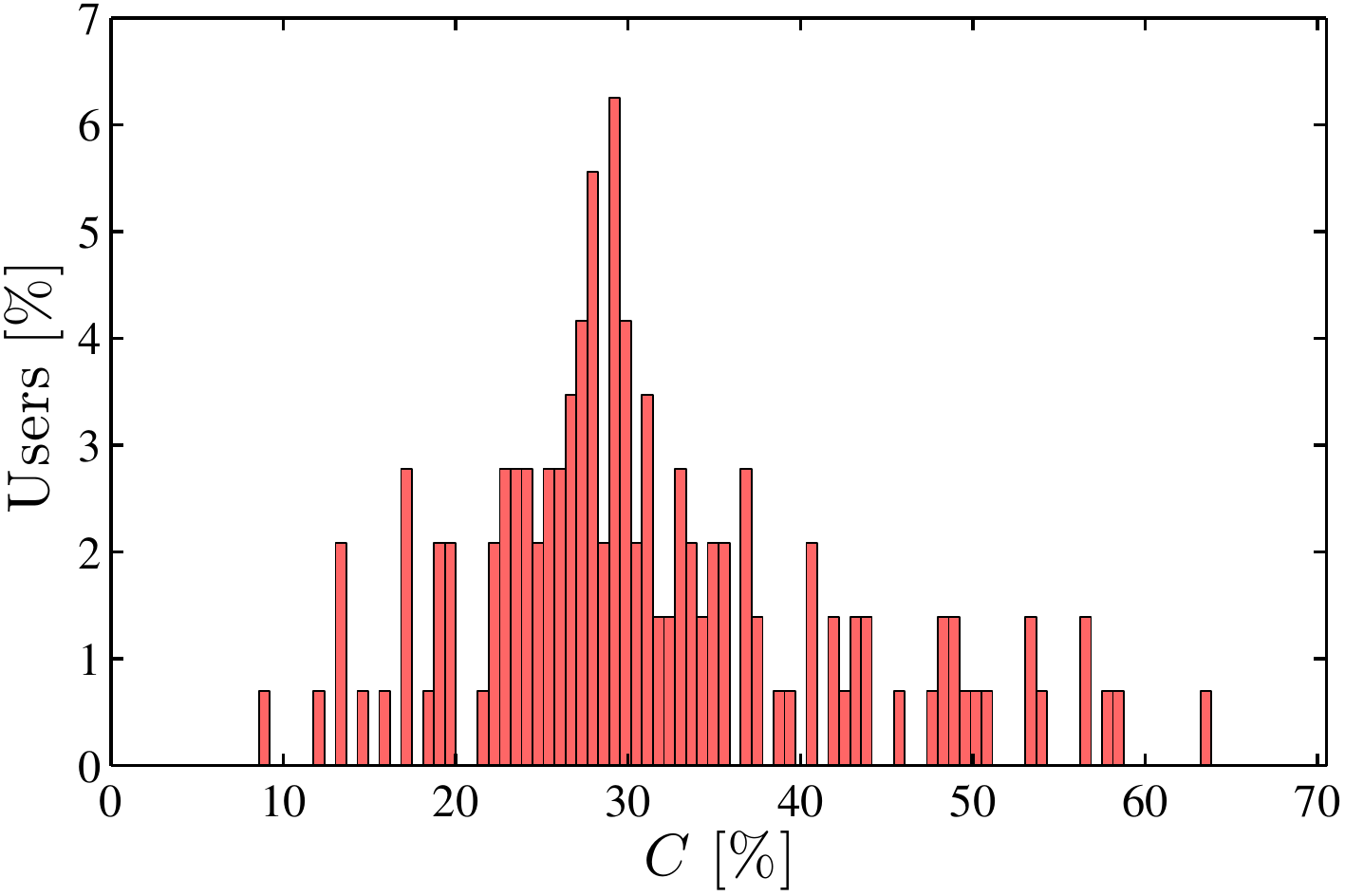}%
}\hspace*{\fill}
\caption{PMF of the expected delay (a) and the buffer capacity (b) when all users apply a deferral rate $\varphi = \phicrit$. The results for (b) are relative to the total number of messages posted by those users.}
\label{fig11}
\end{figure*}

The following two figures, Fig.~\ref{fig11}(a) and Fig.~\ref{fig11}(b), show the PMF of the expected delay and the buffer capacity, in the case when all users wish to attain the critical privacy, i.e., when they apply their corresponding critical rates. The presented results were obtained analytically by using the expressions derived in Sec.~\ref{sec:DelayCapacity:Theory}. From Fig.~\ref{fig11}(a), we check that the minimum, mean and maximum observed values for $\bar{\delta}|_{\varphi=\phicrit}$ are 1.18, 3.89 and 9.05 hours, respectively. As for the buffer capacity, Fig.~\ref{fig11}(b) shows that the minimum, mean and maximum observed values for $C|_{\varphi=\phicrit}$ are 8.92, 31.24 and 63.52\% of users' messages.

\begin{figure}[t]%
\centering
\includegraphics[scale=\SinglePlotScale]{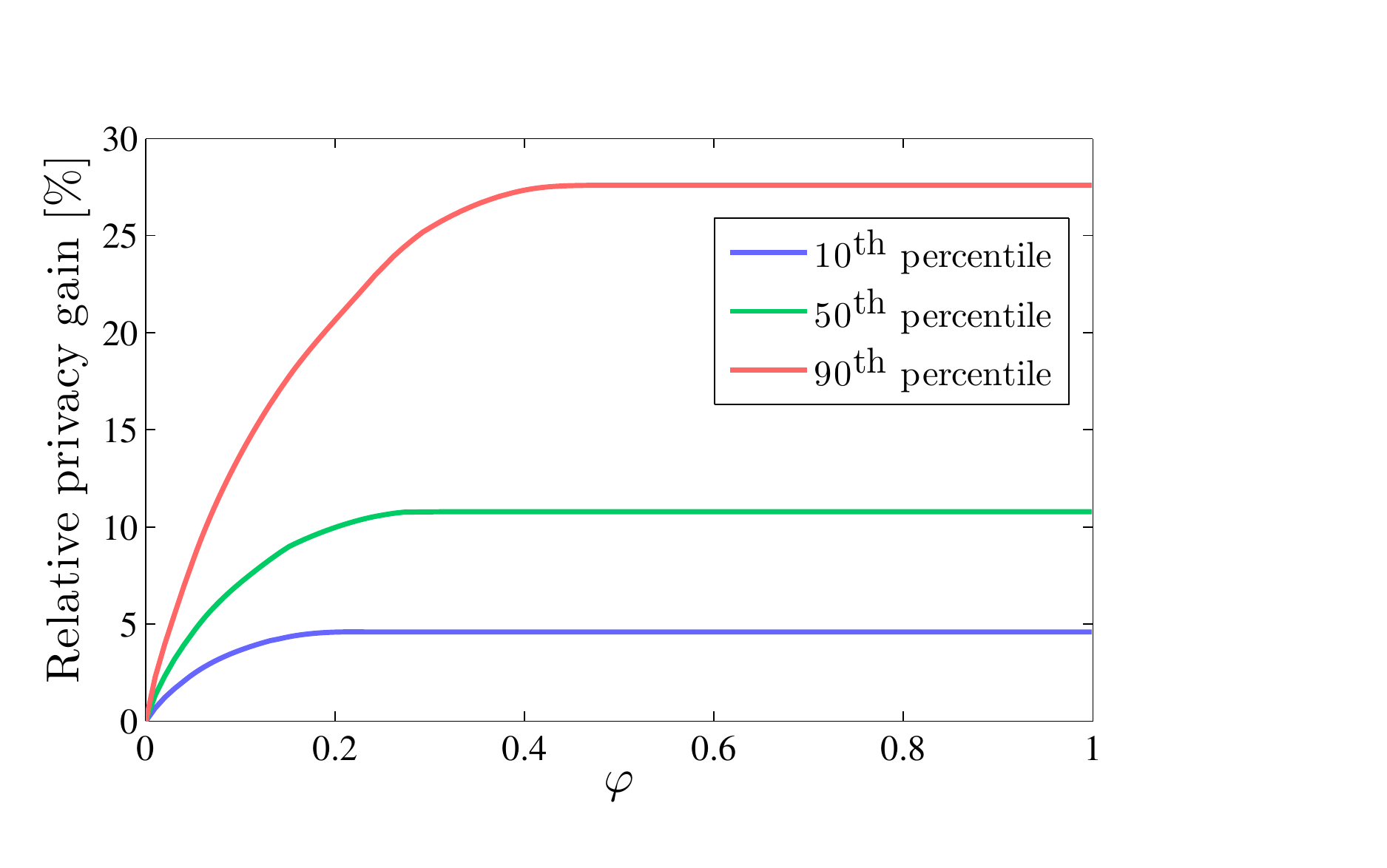}
\caption{Percentile curves of relative privacy gain for different values of $\varphi$.}
\label{fig12}
\end{figure}

The second set of experiments contemplates a scenario where all users apply our privacy-enhancing mechanism by using a common message-deferral rate. Under this assumption, Fig.~\ref{fig12} shows the privacy protection achieved by those users in terms of percentile curves (10\textsuperscript{th}, 50\textsuperscript{th} and 90\textsuperscript{th}) of relative privacy gain. These results were obtained by applying the closed-form expression for the optimal storing and forwarding strategies derived in~\cite[\S 7]{Parra13PhD}. Specifically, we computed the optimal strategies of each user for 100 uniformly distributed values of $\varphi \in[0,0.999]$. However, because a user would not apply rates beyond $\phicrit$, our evaluation uses the solution to the problem $\cP(\phicrit)$ when $\varphi >\phicrit$.

In this figure, we observe how the percentile curves of relative privacy gain increase with $\varphi$ until a certain rate, beyond which these curves are constant. This is consistent with the fact that users attain the maximum level of privacy,  $\log n$, for $\varphi \geqslant \phicrit$. An interesting conclusion that can be drawn from Fig.~\ref{fig12} is that users in our data set will require relatively small margins of privacy gain to achieve the critical-privacy level. This may be observed, for example, for $\varphi=0.60$, i.e., when almost all users get their maximum level of privacy, according to Fig.~\ref{fig10}. Concretely, for this value of $\varphi$, the 10\textsuperscript{th}, 50\textsuperscript{th} and 90\textsuperscript{th} percentile curves show privacy gains of only 4.59\%, 10.78\% and 27.60\%, respectively.

Our last series of experiments analyze the impact of our mechanism from the point of view of message traffic load. Recall that the objective of message deferral is to maximize the Shannon entropy of the apparent profile and thus to spread user activity uniformly over time. This is obviously beneficial from the standpoint of user privacy, according to Jaynes' rationale. But at the same time, entropy maximization may help social networking sites manage their networking resources more efficiently, as our mechanism contributes to distribute the traffic load evenly.

\begin{figure*}[t]%
\centering\hspace*{\fill}
\subfigure[$\varphi \simeq 0.1221$.]%
{\includegraphics[scale=\FigScaleMultipleFourB]{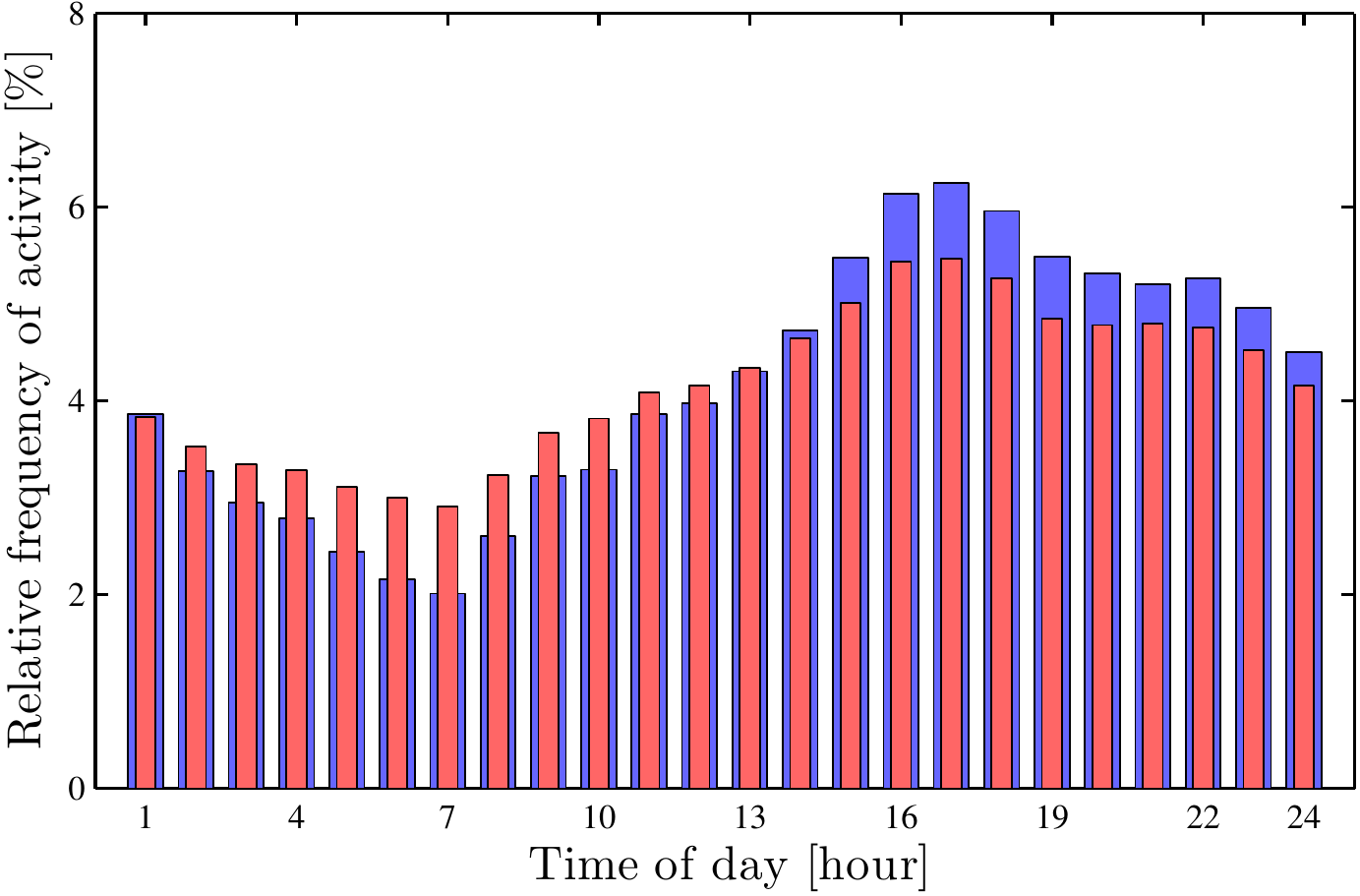}}%
\hfill
\subfigure[$\varphi \simeq 0.2422$.]%
{\includegraphics[scale=\FigScaleMultipleFourB]{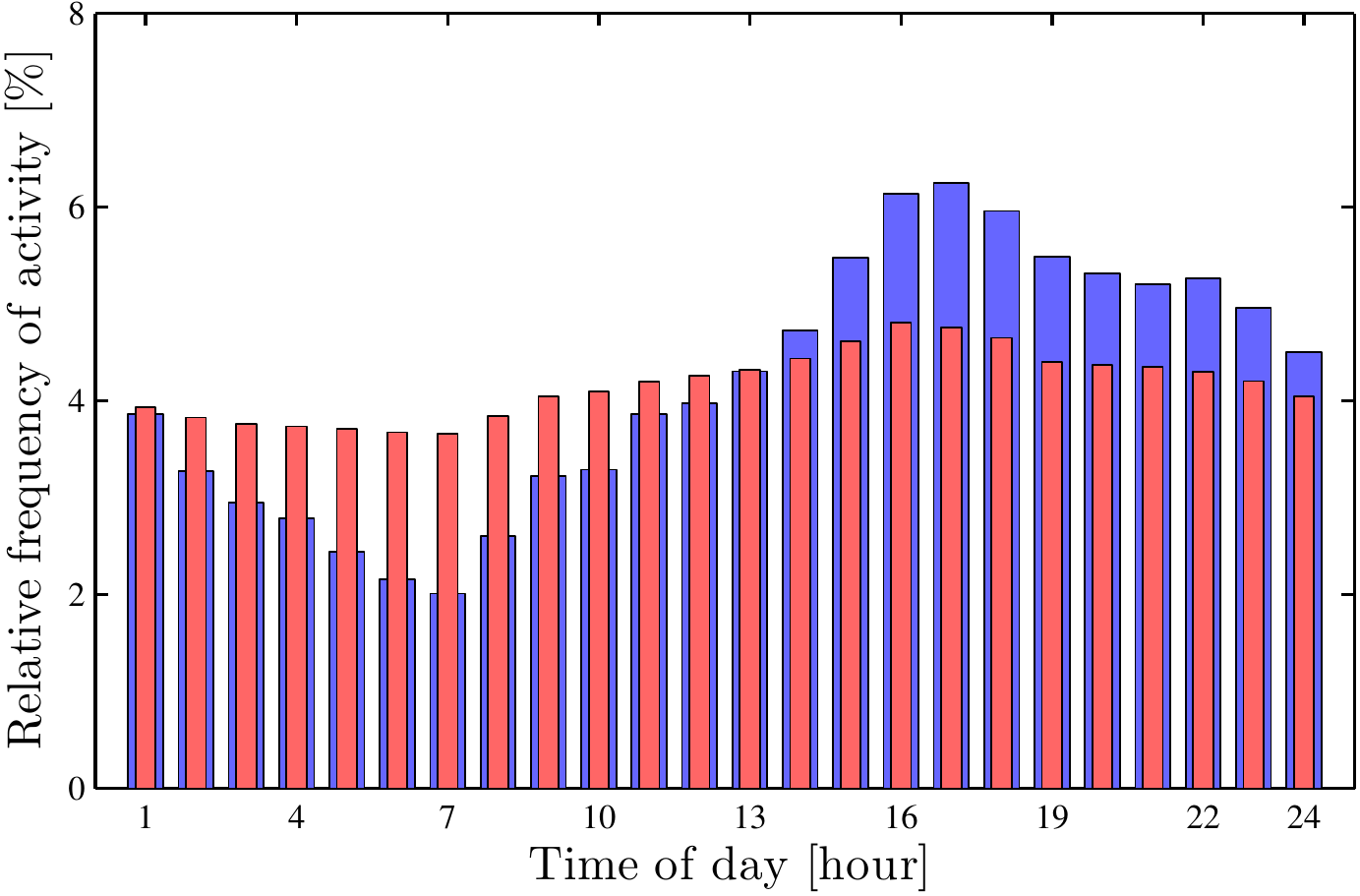}}%
\hspace*{\fill}
\\
\hspace*{\fill}
\subfigure[$\varphi \simeq 0.3633$.]%
{\includegraphics[scale=\FigScaleMultipleFourB]{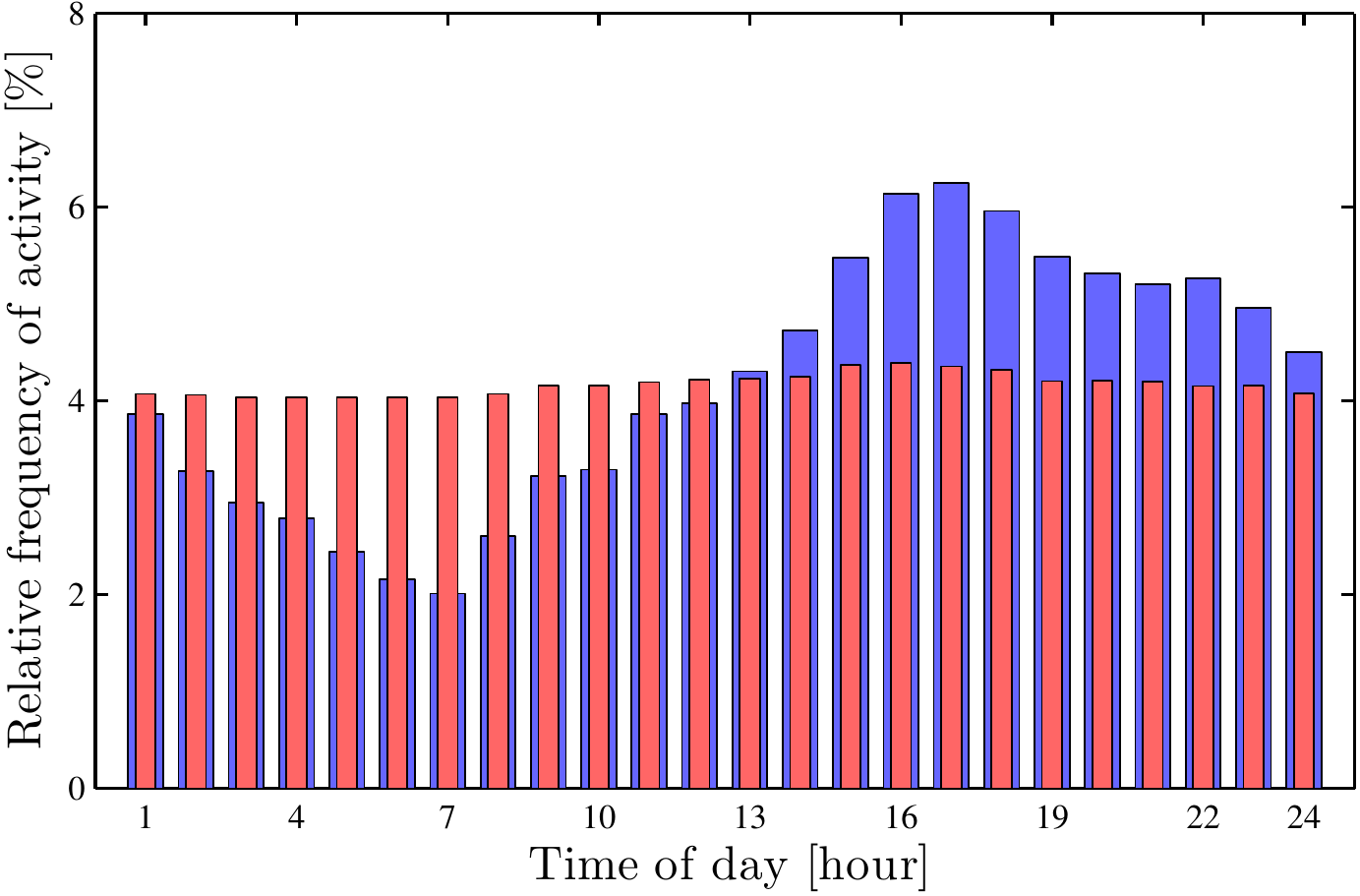}}%
\hfill
\subfigure[$\varphi \simeq 0.4844$.]%
{\includegraphics[scale=\FigScaleMultipleFourB]{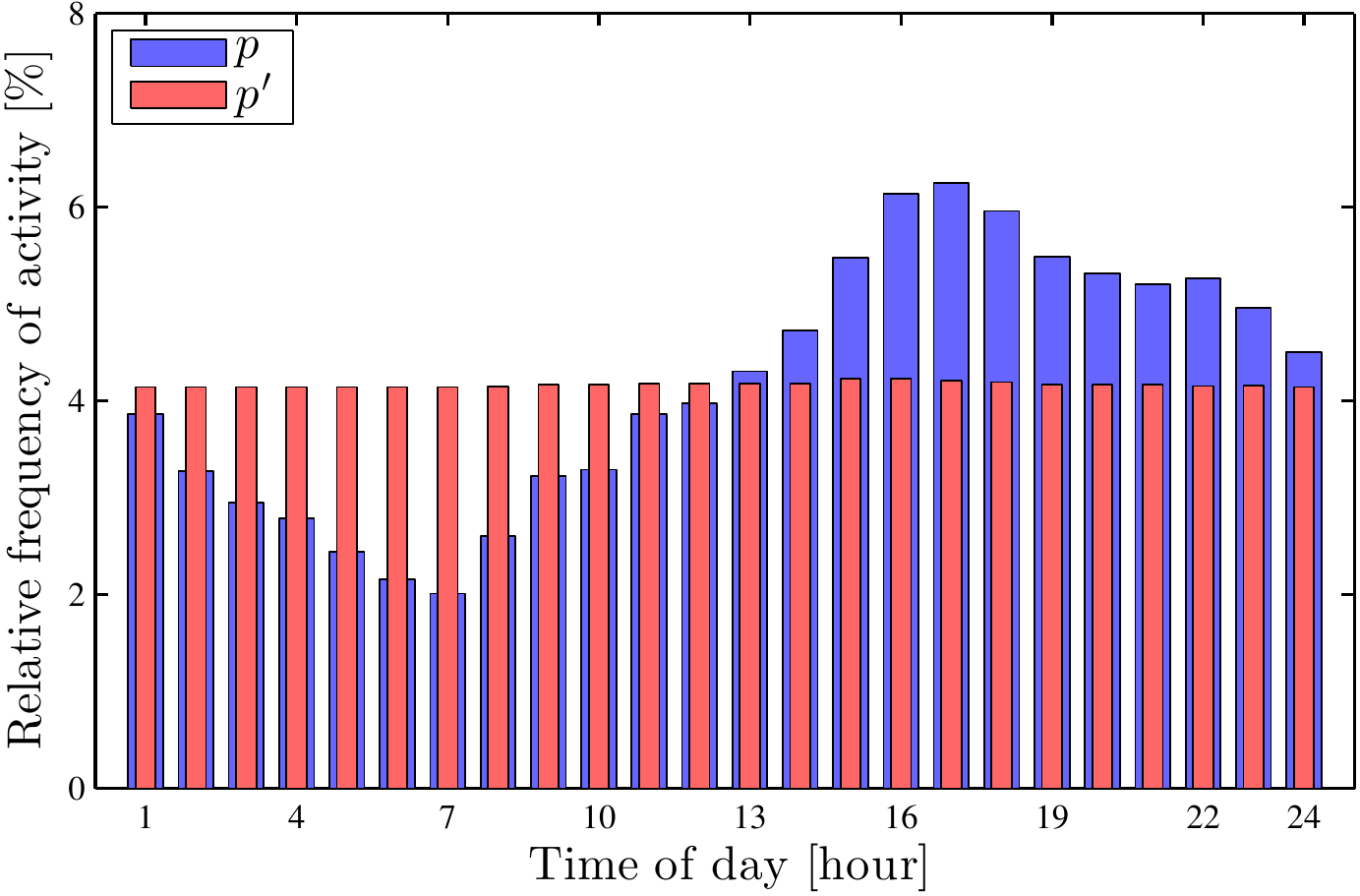}}%
\hspace*{\fill}
\caption{Relative histogram of the tweets in our data set within one day. We denote this histogram as $p$. As a consequence of the deferral of those tweets, the profile $p$ results in the modified profile $p'$.}
\label{fig13}
\end{figure*}

Fig.~\ref{fig13} illustrates this point. In particular, it shows the percentage of messages posted to Twitter by our set of users within a day. Since we computed this as the aggregated profile of all users, we refer to it as the \emph{population}'s profile $p$. The modified version of this relative histogram due to our mechanism is denoted by $p'$. We have represented this profile by assuming that all users apply a common message-deferral rate.

Not entirely unexpectedly, Fig.~\ref{fig13}(a) shows that the time slots most affected by our PET are those with the lowest and highest activity. This is the case of the intervals 5, 6, 7 and 8 on the one hand, and 15, 16, 17, 18 and 19 on the other. For this relatively small value of deferral rate, the number of messages posted between 6 a.m.\ and 7 a.m.\ is increased by 44.68\%, whereas the amount of messages sent between 16 p.m.\ and 17 p.m.\ is reduced by 12.50\%. In Fig.~\ref{fig13}(d), $\varphi\simeq0.4844$ and the overall profile of activity $p'$ becomes nearly uniform. In this last case, the largest increase in the number of tweets is observed for the time slot 7, while the largest reduction in the number of tweets is spotted for the time period 17. In particular, in those time intervals we observe an increase and a reduction of 106.03\% and 32.65\%, respectively. In summary, should our data set be representative of the whole population of Twitter users, the extensive application of the proposed PET could reduce substantially the number of networking resources and maximize the efficiency of such resources.

\section{Concluding Remarks}
\label{sec:Conclusion}
\noindent
Motivated by the lack of previous works specifically addressing the threat of time profiling in social networks, as well as the danger that such type of attack entails, the paper at hand presents a novel, smart message-deferral mechanism. This approach consists in an intelligent delay of a given number of messages posted by users in social networks in a manner that the observed profiles generated by the attacker do not break the privacy of those users. In other words, the attacker is unable to infer any time-based sensitive information by just observing and logging the timestamp of each interaction of the end users with the social networking sites.

Moreover, a detailed architecture implementing this mechanism has been described and analyzed, showing the feasibility of our proposal. Yet, any PET comes at the cost of certain utility loss. Hence, we have studied two meaningful utility metrics specific for our smart deferral mechanism (both in terms of the message deferral rate), namely: expected message delay and messages storage capacity. As shown, both metrics exhibit an increasing, nonlinear behavior with regards to the deferral rate. When the critical deferral rate (beyond which the maximum level of privacy is attained) is known, those outcomes become remarkably helpful to assess the optimal capacity for the messages buffer, as well as the average expected delay of each message in the system.

Finally, a comprehensive set of experiments has been conducted (analyzing the behavior of 144 Twitter users), demonstrating the suitability and accuracy of our solution. In particular, it has been proved that most of the studied users will not require delaying a large percentage of their tweets for their apparent profiles to become the uniform distribution. Likewise, users in our data set will require relatively small margins of privacy gain to achieve the critical-privacy level. Another interesting conclusion states that our approach may help social networking sites manage their networking resources more efficiently, as it contributes to distribute the traffic load evenly. Furthermore, the mean values for the messages expected delay and messages storage capacity in our experiments, respectively, was 3.89 hours and 31.24\% of users' messages

As for the future research lines derived from this work, we are investigating some of the assumptions made in this work. Thus for instance, since we acknowledge that the user activity may vary significantly over time, we need to consider this fact in order to periodically update users' profiles. In the same direction, we want to study the bootstrapping problem, i.e., how to define users' profiles when the system is launched for the first time, or while the system is learning the actual users' profiles.

\ifCLASSOPTIONcompsoc
  \section*{Acknowledgments}
  We would like to express our gratitude to Silvia Puglisi for retrieving the data set used in the experimental section. This work was partly supported by the Spanish Government through projects Consolider Ingenio 2010 CSD2007-00004 ``ARES'', TEC2010-20572-C02-02 ``Consequence'' and by the Government of Catalonia under grant 2009 SGR 1362.
\else
  \section*{Acknowledgment}
\fi


\end{document}